\documentclass[journal]{IEEEtran}  

\usepackage{graphicx,des,ieee_thm,amsmath,psfrag,url}     
\usepackage{amsmath,array,cite,tikz}
\usetikzlibrary{shapes,arrows}
\usepackage{tikz-cd}
\usepackage{graphics} 
\usepackage{epsfig} 
\usepackage{mathptmx} 
\usepackage{enumerate}
\usepackage{mathrsfs}
\usepackage{float}
\usepackage{array}
\usepackage{pstricks}
\usepackage{pst-node}
\usepackage{pst-blur}
\usepackage{tikz}
\usepackage{xhfill}
\usepackage{lipsum}
\usetikzlibrary{fit,positioning,backgrounds}
\usetikzlibrary{positioning}
\usetikzlibrary{arrows, decorations.markings}
\usetikzlibrary{arrows.meta, bending, positioning, shapes}
\definecolor{Pink}{rgb}{1.,0.75,0.8}

\IEEEoverridecommandlockouts                              

\newtheorem{remark}{Remark}

\title{\LARGE \bf
	Compositional and Abstraction-Based Approach for Synthesis of Edit Functions for Opacity Enforcement}

\author{Sahar Mohajerani, Yiding Ji and St\'ephane Lafortune
	\thanks{The work of the first author was supported by the Swedish Research Council. 
		The work of the second and third authors was supported in part by US NSF grants CNS-1421122 and CNS-1738103.}
	\thanks{Sahar Mohajerani, Yiding Ji and St\'ephane Lafortune are with the Department of Electrical Engineering and Computer Science, University of Michigan, Ann Arbor, MI, USA. {\tt\small\{saharm;jiyiding;stephane@umich.edu\}}}%
}

\begin{document}
	\newcommand*{\QEDA}{\hfill\ensuremath{\blacksquare}}%
	\newcommand*{\QEDB}{\hfill\ensuremath{\square}}%
	\newcommand{\calr}{\mathcal{R}}
	\newcommand{\supcn}{\ensuremath{\mathrm{sup}\mathcal{C}}}
	\newcommand{\ACTE}{\Sigma_\tau}
	
	\newcommand{\hdet}{H_{b}}
	\newcommand{\hdes}{H_{obd}}
	\def\figscale{.43}
	\maketitle

	\tikzstyle{dim} = [draw, diamond, aspect=2, text width=4.2em, text centered, node distance=1cm, inner sep=0pt]
	\tikzstyle{dimd} = [draw, diamond, dashed,aspect=2, text width=4.2em, text centered, node distance=1cm, inner sep=0pt]
	\tikzstyle{dimdo} = [draw, diamond, dotted,aspect=2, text width=4.2em, text centered, node distance=1cm, inner sep=0pt]
	\tikzstyle{ini} = [draw=none,fill=none]
	
	\tikzstyle{sum} = [circle, scale=0.3, node distance=4cm, inner sep=0pt,draw,fill]
	\tikzstyle{dot} = [circle, scale=1, node distance=4cm, inner sep=0pt,draw,fill]
	\tikzstyle{rec} = [rectangle, draw, text centered, node distance=1cm, minimum height=2em]
	\tikzstyle{recd} = [rectangle, draw, dashed,text centered, node distance=1cm, minimum height=2em]
	\tikzstyle{recdo} = [rectangle, draw, dotted,text centered, node distance=1cm, minimum height=2em]
	
	\tikzstyle{inirec} = [rectangle, draw, text centered, node distance=.5cm, minimum height=2em]
	\tikzstyle{line} = [draw, -latex']
	
	\tikzstyle{ov} = [draw, ellipse,  anchor=west,  node distance=1cm, minimum height=0.25em]
	\tikzstyle{ovd} = [draw, ellipse,  dashed,anchor=west,  node distance=1cm, minimum height=0.25em]
	\tikzstyle{ovdo} = [draw, ellipse,  dotted,anchor=west,  node distance=1cm, minimum height=0.25em]
	
	\tikzstyle{dimini} = [draw, diamond, aspect=2,  text centered, node distance=1.5cm, inner sep=0pt]
	\tikzstyle{dimtrans} = [draw, diamond, aspect=2, text centered, node distance=1cm, inner sep=0pt]
	\tikzstyle{inov} = [draw, ellipse, anchor=west,  node distance=2.3cm, minimum height=0.25em]
	\tikzstyle{dashed}=                  [dash pattern=on 3pt off 3pt]
	\tikzstyle{vecArrow} = [thin, decoration={markings,mark=at position
		1 with {\arrow[thin]{open triangle 60}}},
	double distance=1pt, shorten >= 4.5pt,
	preaction = {decorate},
	postaction = {draw,line width=0.4pt, white,shorten >= .5pt}]
	\tikzstyle{innerWhite} = [thin, white,line width=1.4pt, shorten >= 4.5pt]
	
	%
	
	\begin{abstract} 
	This paper develops a novel compositional and abstraction-based approach to synthesize edit functions for opacity enforcement in modular discrete event systems. 
	Edit functions alter the output of the system by erasing or inserting events in order to obfuscate the outside intruder, whose goal is to infer the secrets of the system from its observation. 
	We synthesize edit functions to solve the opacity enforcement problem in a modular setting, which significantly reduces the computational complexity compared with the monolithic approach.  
	Two abstraction methods called opaque observation equivalence and opaque bisimulation are first employed to abstract the individual components of the modular system and their observers. Subsequently, we propose a method to transform the synthesis of edit functions to the calculation of modular supremal nonblocking supervisors. We show that the edit functions synthesized in this manner correctly solve the opacity enforcement problem. 
		
	\end{abstract}

	\section{INTRODUCTION}

	Opacity characterizes whether the integrity of the secrets of a system can be preserved from the inference of an outside intruder, potentially with malicious purposes. The intruder is modeled as a passive observer with knowledge of the system's structure. A system is called opaque if the intruder is unable to infer the system's secrets from its observation.

	Starting with~\cite{bryans2005modelling, bryans2008opacity} in the computer science literature, opacity has been extensively studied, especially in the field of discrete event systems (DES), under multiple frameworks. 
	
	For finite state automaton models, various notions of opacity have been studied, e.g., language-based opacity~\cite{lin2011opacity}, current-state opacity~\cite{saboori2007notions}, initial-state opacity~\cite{saboori2013verification}, K-step opacity~\cite{yin2017new} and infinite-step opacity~\cite{saboori2012verification}. 
	Opacity has also been discussed in some other system models, like infinite state systems~\cite{chedor2015diagnosis}, modular systems~\cite{masopust2019complexity} and Petri nets~\cite{tong2017verification, tong2017decidability}. Opacity under a special observer called Orwellian observer is discussed in~\cite{mullins2014opacity} and opacity under powerful attackers is studied in~\cite{helouet2018opacity}. A more recent work~\cite{yin2018verification} investigates opacity for networked supervisory control systems. 
	Furthermore, some works investigate opacity in stochastic settings, e.g., ~\cite{keroglou2017probabilistic, berard2015probabilistic, chen2017quantification, wu2018privacy}. Specifically, \cite{yin2019infinite} presents a novel approach to tackle infinite-step and K-step opacity in stochastic DES. 
	The survey paper~\cite{jacob2016overview} summarizes some recent results on opacity in DES. 
	
	When opacity does not hold, it is natural to study its enforcement~\cite{falcone2015enforcement}.
	One popular approach is supervisory control~\cite{dubreil2010supervisory, darondeau2015enforcing, takai2008formula, saboori2012opacity, yin2016uniform}, where some behaviors of the system are disabled before they reveal the secrets. 
	Another widely-applied method is sensor activation~\cite{cassez2012synthesis, zhang2015maximum, yin2019general}, where the observability of events is dynamically changed. 	
	
	Recently, a new enforcement method called insertion function was proposed in~\cite{wu2014synthesis}, which inserts fictitious events into the output of the system to obfuscate the intruder. 
	The authors of~\cite{ji2018enforcement} extended the method to study opacity enforcement under the assumption that the intruder may or may not know the implementation of insertion functions, while~\cite{ji2019enforcing} discussed opacity enforcement by insertion functions under quantitative constraints. 
	As a following work, \cite{wu2018synthesis} investigates a more general method called edit functions, which manipulate the output of the system by either inserting or erasing events. Then~\cite{ji2017edit, ji2019synthesis} considers the case when the edit function's implementation is known to the intruder. 
	As a summary and extension, \cite{ji2019synthesis} characterizes opacity enforcement by edit functions as a three-player game and proposes a novel information structure called three-player observer (TPO) to embed edit functions. A special TPO called the All Edit Structure (AES) is also introduced in~\cite{ji2019synthesis} to characterize the edit constraints. 

	In this work, we elaborate the method in~\cite{ji2019synthesis} to study opacity enforcement in a modular setting. Our motivation is as follows. 
	To generate a three-player observer, the observer of the system needs to be calculated, which is potentially costly in computation. Furthermore, modern engineering systems usually contain multiple components that are synchronized and subject to malicious inference. In this sense, if we are to apply edit functions to enforce opacity, heavy computation is involved both from determining individual systems and synchronizing them, which may be potentially cumbersome.

	To alleviate this issue, this paper applies a compositional and abstraction-based method to reduce the size of the modular system before calculating the All Edit Structure. \emph{Bisimulation} and \emph{observation equivalence}~\cite{milner1989communication} are well-known methods to abstract the state space of an automaton, while they do not preserve opacity properties in general. As a variant, \cite{zhang2018opacity} proposes several innovative concepts termed opacity-preserving (bi)simulation relations to reduce the state space of the system in opacity verification. A compositional visible bisimulation equivalence method is discussed in ~\cite{noori2018compositional} for abstraction-based opacity verification. 

	For abstraction, we introduce \emph{opaque observation equivalence} and \emph{opaque bisimulation}, which consider the secrecy status of states when merging them. In our framework, each individual system is abstracted using opaque observation equivalence. After that, the observer is calculated. Since abstraction reduces the size of the state space, the computational complexity of calculating the observer is lowered potentially. Next, opaque bisimulation is employed to the observer of each abstracted individual system, resulting in the smallest possible automaton for future discussion. 

	We further leverage some results from supervisor reduction and modular supervisory control theory to reduce the complexity of supervisor synthesis. 
	There is a rich literature on both topics, see, e.g., ~\cite{su2004supervisor, su2010model, schmidt2012efficient, malik2013compositional, mohajerani2014framework}. The main idea is to convert the construction of the monolithic All Edit Structure to a modular supervisory control problem. Specifically, we first transfer each individual three-player observer (without considering edit constraints) to its automaton form and view the set of interacting automata as the ``plant" to be controlled. Then we put the edit constraint as the specification, also in an automaton form. Afterwards, we perform modular supervisory control to synthesize a least restrictive and nonblocking modular supervisor. It is shown that all the traces accepted by the supervisor represent valid edit decisions contained in the monolithic AES. Compared with the conventional monolithic approach for supervisor synthesis~\cite{cassandras2009introduction}, our compositional approach is more efficient in computation.

	The presentation of this work is organized as follows.
	Section~\ref{sec:preli} gives a brief background introduction about the system model, supervisory control theory and edit functions. The general idea of the paper is presented in \Sect~\ref{sec:generalIdea} as a flow chart.
	Section~\ref{sec:AES} explains the abstraction methods and synchronization of three-player observers.
	Next, Section~\ref{sec:sup} transforms the calculation of the monolithic All Edit Structure to the calculation of a modular supervisor.
	Finally, some concluding remarks are given in Section~\ref{sec:conclusion}. 
	
	A preliminary and partial version of this work appears in~\cite{ji2018efficient}. The current work improves~\cite{ji2018efficient} in the sense that \cite{ji2018efficient} only considers abstraction methods to synthesize edit functions in a monolithic setting, while this work also takes synchronous composition into consideration and the edit functions are synthesized by a modular approach. 
	\def\figscale{.1}
	
	\section{MODELING FORMALISM AND BACKGROUND}\label{sec:preli} 

	\subsection{Events, Automata and their Composition}
	
	In this work, we consider discrete  event systems modeled as deterministic or nondeterministic automata. 
	
	\begin{definition} 
		A (nondeterministic) finite-state automaton is a tuple $G= \langle \Sigma, Q, \rightarrow, Q^0\rangle$,
		where $\Sigma_{}$ is a finite set of events, $Q$ is a finite set of states,
		$\intrans \subseteq Q \times \Sigma_{} \times Q$ is the \emph{state
			transition relation}, and $Q^0 \subseteq Q$ is the set of
		\emph{initial states}.
		$G$~is \emph{deterministic} if $|Q^0| = 1$ and if $x
		\trans[\sigma] y_1$ and $x \trans[\sigma] y_2$ always implies that $y_1 = y_2$.
	\end{definition}
	
	When state marking is considered, the above definition is extended to $G=\langle \Sigma, Q, \trans,Q^0, Q^m\rangle$, where $Q^m\subseteq Q$ is the set of \emph{marked states}.  
	In this paper, we identify marked states using gray shading in the figures.

	We assume that the system is partially observed, thus the concepts of \emph{observable} and \emph{unobservable} events are introduced. Since the exact identity of unobservable events is irrelevant in our later discussion of opacity, they are uniformly represented by a special event $\tau$. The event $\tau$ is never included in the alphabet $\ACT$, unless explicitly mentioned. For this reason, $\ACT_\tau=\ACT\cup\{\tau\}$ is used to represent the whole set of observable and unobservable events. 
	Hereafter, nondeterministic automata may contain transitions labeled by $\tau$, while \emph{deterministic} automata \emph{never} contain $\tau$ transitions. Moreover,  $P_\tau:\ACT_\tau^*\to \ACT^*$ is the \emph{projection}
	that removes from strings in $\ACT_\tau^*$ all the $\tau$ events.
	
	When automata are brought together to interact, lock-step synchronization
	in the style of~\cite{hoare1978communicating} is used.
	\begin{definition}
		\label{def:synch}
		Let $G_1 =\langle\ACT_{1}, Q_1, \trans_1, Q^0_1, Q^m_1\rangle$ and $G_2 = \langle\ACT_{2}, Q_2, \trans_2, Q^0_2, Q^m_2\rangle$ be two nondeterministic automata.
		The \emph{synchronous composition} of $G_1$ and $G_2$ is defined as
		\begin{equation}
		\begin{split}
		G_1 \sync  G_2:= \left\langle \Sigma_{1}\cup \Sigma_{2}, Q_1\times Q_2,
		\intrans,
		Q^0_1 \times Q^0_2,
		Q_1^m \times Q_2^m
		\right\rangle 
		\end{split}
		\end{equation}
		where
		$$
		\begin{array}{@{}r@{\quad}l@{}}
		(x_1,x_2) \trans[\sigma] (y_1,y_2) &
		\mbox{if } \sigma \in ({\ACT_1} \cap {\ACT_2}),\\
		& x_1 \trans[\sigma]_1 y_1,\ \text{and}\
		x_2 \trans[\sigma]_2 y_2 \, ; \\
		(x_1,x_2) \trans[\sigma] (y_1,x_2) &
		\mbox{if } \sigma \in (\ACT_1 \setminus \ACT_2)\ 
		\cup\{\tau\}\ \\ & \text{and}\ 
		x_1 \trans[\sigma]_1 y_1 \, ; \\
		(x_1,x_2) \trans[\sigma] (x_1,y_2) &
		\mbox{if } \sigma \in (\ACT_2 \setminus \ACT_1)\ 
		\cup\{\tau\}\ \\&  
		\text{and}\ x_2 \trans[\sigma]_2 y_2 \, .
		\end{array}
		$$

	\end{definition}
	
	Importantly, synchronous composition only imposes lock-step synchronization on common events from $\ACT_1$ and $\ACT_2$. 
	

	
	%

	The transition relation of an automaton $G$ is written in infix notation $x \trans[\sigma] y$,
	and it is extended to strings in $\Sigma_\tau^*$ by letting $x \trans[\varepsilon]
	x$ for all $x\in Q$, and $x\stackrel{t\sigma}{\rightarrow}z$ if
	$x\stackrel{t}{\rightarrow}y$ and $y\stackrel{\sigma}{\rightarrow}z$ for
	some~$y \in Q$. Furthermore, $x \trans[t]$ means that $x \trans[t] y$ for
	some $y \in Q$, and $x \trans y$ means that $x \trans[t] y$ for some $t \in
	\ACT_\tau^*$. These notations also apply to state sets, where $X \trans[t]Y$ for $X, Y
	\subseteq Q$ means that $x \trans[t]y$ for some $x \in X$ and $y\in Y$, and to automata, where $G \trans[t]$ means that $Q^0 \trans[t]$ ($t$ is defined in $G$) and $G \trans[t]x$ means $Q^0 \trans[t]x$. 
	
	For brevity, $p\ttrans[s]q$ for $s\in\ACT^*$ represents the existence of a string $t\in\ACT_{\tau}^*$ such that $P_\tau(t)=s$ and $p\trans[t]q$. Thus, $q\trans[u]p$ for $u\in\ACT_\tau^*$ means a path containing exactly the events in $u$, while $q\ttrans[u]p$ for $u\in\ACT^*$ means existence of a path between $p$ and $q$ with an arbitrary number of $\tau$ events between the observable events in $u$. Similarly, $p\ttrans[\tau] q$ means the existence of a string $t\in \{\tau\}^*$ such that $p\trans[t]q$.  
	
	The \emph{language} of an automaton~$G$
	is defined as $\LANG(G) = \{\, s \in \ACTstar \mid G \ttrans[s] \,\}$ and the language generated by $G$ from $q\in Q$ is $\LANG(G,q)=\{s\in\ACT^*\ |\ q\ttrans[s]\}$, thus we do not include event $\tau$ in the language of an automaton. 
	 Moreover, we also introduce projections $P_i$ for $i=1,2$, which are  $P_i:(\ACT_{1}\cup\ACT_{2})^*\to \ACT^*_i$ for $i=1,2$.
	 

	
	
	For a nondeterministic automaton $G= \langle \ACTE, Q, \rightarrow, Q^0\rangle$, the set of \emph{unobservably reached states} of $B\in 2^Q$, is $UR(B)=\bigcup\{C\subseteq Q\ |\ B\ttrans[\tau] C \}$. Its \emph{observer} $det(G)=\langle \ACT, X_{obs},\trans_{obs},X_{obs}^0\rangle$ is a deterministic automaton, where $X_{obs}^0=UR(Q^0)$ and $X_{obs}\subseteq 2^Q$, and $X\trans[\sigma]_{obs}Y$, where $X, Y\in X_{obs} $, if and only if $Y=\bigcup\{UR(y)\ | \ x\trans[\sigma] y\ \textnormal{for some}\ x\in X \ \textnormal{and}\ y\in Q \}$. By convention, only reachable states from $X^0_{obs}$ under $\trans_{obs}$ are considered in this paper. We also refer to the observer as the \emph{(current-state) estimator} of the system while an observer state is referred to as \emph{(current-state) estimate}. 
	
	A common automaton operation is the \emph{quotient} modulo, which is an
	equivalence relation on sets of states.
	
	\begin{definition}
		Let $Z$ be a set. A  relation $\insim \subseteq  Z\times Z$ is called an
		\emph{equivalence relation} on~$Z$ if it is reflexive,
		symmetric, and transitive.
		Given an equivalence relation~\insim\ on~$Z$, the \emph{equivalence class} of $z \in Z$
		is $[z] = \{\, z' \in Z \mid z \sim z' \,\}$,
		and $\tilde{Z} = \{\, [z]
		\mid z \in Z \,\}$ is the set of all equivalence classes modulo~$\sim$.
	\end{definition}
	
	\begin{definition}\label{def:quotient}
		Let $G = \langle \Sigma, Q, \rightarrow, Q^0\rangle$ be an automaton and let $\insim \subseteq Q \times Q$ be an equivalence relation. The \emph{quotient automaton} of $G$
		modulo~$\sim$ is $
		\tilde{G} = \langle \ACT, \tilde{Q}, \intrans\modsim, \tilde{Q}^0\rangle$, 
		where $\intrans\modsim =\{\, ([x],\sigma,[y]) \mid x \trans[\sigma] y
		\,\}$ and $\tilde{Q}^0 =\{\, [x^0] \mid x^0 \in Q^0 \,\}$.
	\end{definition}

	In order to compare automata structurally, we say that an automaton is a \emph{subautomaton} of another automaton if all states and transitions in the first automaton are contained in the second one. 
	Formally, we have the following definition:
	\begin{definition}
		
		Let $G_1=\left\langle \Sigma_{\tau},Q_1,\rightarrow_1,Q^0_1,Q^m_1\right\rangle$ and $G_2=\left\langle\Sigma_{\tau},Q_2,\rightarrow_2, Q^0_2,Q^m_2\right\rangle$ be two automata. 
		$G_1$ is a \textit{subautomaton} of $G_2$, denoted by $G_1 \sqsubseteq G_2$, if $Q_1 \subseteq Q_2$, $\mathord{\rightarrow}_1 \subseteq \mathord{\rightarrow}_2$, $Q^0_1 \subseteq Q^0_2$, and
		$Q^m_1 \subseteq Q^m_2$.
	\end{definition}

	\subsection{Supervisory Control Theory} 
	\label{sec:sct}

Considering plant $G$ and specification $K$, \textit{supervisory control theory} provides a method to synthesize a supervisor to restrict the behavior of the plant such that the given specification is always fulfilled. The supervisor $S$ is a function defined from the language of the system $G$ to the set of events, formally, $S: \mathcal L(G)\rightarrow 2^{\Sigma}$.
	We also partition the set of events as \emph{uncontrollable events} and \emph{controllable events}, i.e., $\Sigma=\Sigma_{uc}\cup \Sigma_c$, where uncontrollable events cannot be disabled by the supervisor. In the figures the uncontrollable events are marked by an exclamation mark (!).  The readers may refer to~\cite{cassandras2009introduction} for the main results of monolithic supervisory control under full observation. Here we focus on concepts and definitions relevant to the present paper  and the synthesis procedure in this paper is done on  deterministic automata.
	Two requirements for the supervisor are \textit{controllability} and \textit{nonblockingness}, where controllability captures \emph{safety} in the presence of uncontrollable events and nonblockingness focuses on \emph{liveness} of the system. 
	
	\begin{definition}
		\label{defNondetControllability}
		\cite{cassandras2009introduction} Let $G = \left\langle \Sigma,Q_G,\rightarrow_G,Q^0_G,Q^m_G\right\rangle$ and $K = \left\langle \Sigma,Q_K,\rightarrow_K,Q^0_K,Q^m_K\right\rangle$ be two deterministic automata such that 
		$K \sqsubseteq G$. 
		$K$~is \emph{controllable} w.r.t.~$G$ if, for all states $x \in Q_K$
		and $y \in Q_G$ and for every uncontrollable event
		$\upsilon \in \Sigma_{uc}$ such that $x \stackrel{\upsilon}{\rightarrow}_G y$,
		it also holds that $x \stackrel{\upsilon}{\rightarrow}_K y$.
	\end{definition}
	
	\begin{definition}
		\label{defNonblocking}
		\cite{cassandras2009introduction} Let $G$ be a deterministic automaton. A state $x$ is called \emph{reachable}
		in~$G$ if $G \rightarrow x$, and \emph{coreachable} in~$G$ if $x \rightarrow Q^m$.
		The automaton~$G$ is called reachable or coreachable if
		every state in~$G$ has this property. $G$~is called \emph{nonblocking} if
		every reachable state is coreachable.
	\end{definition}
	
	The upper bound of controllable and nonblocking subautomata is again
	controllable and nonblocking. This implies the existence of a least
	restrictive subautomaton of the original system, which is achieved by the maximally-permissive and nonblocking supervisor.

	\begin{definition}
		Let $G$ be an automaton, the supremal controllable and nonblocking subautomaton of $G$ is called the supremal supervisor, denoted by $\supCN(G)$ where for all controllable and nonblocking automata $K$ w.r.t.\ $G$, $K\sqsubseteq \supCN(G)$.
	\end{definition}
	
	Synthesis of  $\supCN(G)$ is done by iteratively removing blocking and uncontrollable states, until a fixed point is reached, and restricting the final automaton to the remaining states and their associated transitions, for more details  please see~\cite{cassandras2009introduction,flordal2007compositional, WonKai:19}.

	In this paper, we assume that the modular system has a set of interacting components $\{G_1,\ldots, G_n\}$, and there is also a set of supervisors in a modular structure, i.e.,
	$\SYSS=\{S_1,\ldots,S_n\}$. Here supervisor $S_i$ is responsible for controlling $G_i$. The set of modular supervisors may be synchronized as $\bigsync^n_{i=1} S_i$. 
	
	

	\subsection{Opacity and Edit Functions}
	\label{sec:edit-fn}
	

	In this work, we suppose system $G$ has certain secret information which is characterized by the set of states. Thus the state space is partitioned into two disjoint subsets: $Q=Q^S\cup Q^{NS}$ where $Q^S$ is the set of \emph{secret states} capturing the secrets of the system, while $Q^{NS}$ is the set of \emph{non-secret states}.  When the system $\SYSG$ is modular, $\SYSG=\{G_1,\ldots, G_2\}$,  the set of secret states of the system, $Q_S$, is $Q^S= \{(x_1,\ldots,x_n)\ | \exists x_i \in Q_i^S \ \}$.
	
	Suppose there is an external intruder modeled as the observer of the system, which intends to infer the secrets of the system from its observation. Then a system is called \emph{opaque} if the intruder is unable to determine unambiguously if the system has entered a secret state or not.  Different notions of opacity have been introduced in literature and we focus on current-state opacity in this work.
	\begin{definition}\label{def:currentStateOp}
		A nondeterministic automaton $G$ with a set of secret states $Q^S$ is \emph{current-state opaque} w.r.t.\ $Q^S$ if $(\forall s\in\LANG(G,q^0) \ \colon\ Q^0\ttrans[s]Q^S)$ then $[\ Q^0\ttrans[s]Q^{NS}]$.
		
	\end{definition}
	
	The system is current-state opaque if for any string reaching a secret state there is string with the same sequence of observable events reaching a non-secret state. It is known that current-state opacity can be verified by building the standard observer automaton. 
	
	\begin{theorem}\label{prop:currentStateOpEstimator} 
		Let $G=\auttuple{\ACT_{\tau}}$ be a nondeterministic automaton with set of secret states $Q^S$.  
		Let  $det(G) = \langle \ACT,X_{obs},\trans_{obs},X_{obs}^0\rangle$ be the current-state estimator of $G$. Then $G$ is current-state opaque w.r.t.\ $Q^S$ if and only if $[det(G)\trans[s]X \ \text{implies that}\  X\not\subseteq Q^S]$.
	\end{theorem}

	If all states violating current-state opacity are removed from the observer $det(G)$, then the accessible part of the remaining structure is called the \emph{desired observer}, denoted by $det_d(G)=\langle \ACT,X_{obsd},\trans_{obsd},X_{obsd}^0\rangle$. 
	The language generated by the desired observer 
	is referred to as the \emph{safe language}, $L_{\textit{safe}}=\LANG(det_d(G))$. 
	Accordingly, we also define the \emph{unsafe language}, $L_{\textit{unsafe}}=\LANG(G)\setminus L_{\textit{safe}}$.

	If a system is not current state opaque then an interface based approach called~\emph{edit function}~\cite{ji2019synthesis, wu2018synthesis} may be applied to enforce it.
	An edit function may insert events into the output of the system or erase events from the output of the system. It is assumed that the intruder fails to distinguish between an inserted event 
	and its genuine counterpart. 
	Let $\ACT^r=\{\sigma\trans\epsilon\colon \sigma\in \ACT\}$ be the set of ``event erasure'' events.  
	\begin{definition}\label{def:editFunction}
		A \emph{deterministic edit function} is defined as  $f_e: \ACT^* \times \ACT \rightarrow \ACT^*$. Given $s \in \LANG(G)$, $\sigma\in \ACT$, 
		\begin{align*}
		f_e(s, \sigma) =\begin{cases}
		s_I\sigma & ~\mbox{if}~s_I~\mbox{is inserted before}~\sigma\\
		\epsilon & ~\mbox{nothing is inserted and}~\sigma~\mbox{is erased}\\
		s_I&~\mbox{if}~s_I~\mbox{is inserted and}~\sigma~\mbox{is erased}
		\end{cases}
		\end{align*}
	\end{definition}
	
	With an abuse of notation, we also define a string-based edit function $\hat{f}_e$ recursively as: $\hat{f}_e(\epsilon)=\epsilon$, $\hat{f}_e(s\sigma)=\hat{f}_e(s)f_e(s,\sigma)$ for $s\in \Sigma^*$ and $\sigma\in \Sigma$. 
In the sequel, to ease the notational burden, we will drop the~``~$\hat{}$~''~in $\hat{f}_e$ and it will be clear from the argument(s) of $f_e$ which function we are referring to (incremental single-event one or string-based one).

	Two notions termed \emph{public safety} and \emph{private safety} were defined in~\cite{ji2019synthesis} to characterize the behavior of edit functions. In this paper, we consider private safety alone under the assumption that the intruder does not know about the implementation of an edit function.
	\begin{definition}[Private Safety]\label{def:private}
			Given $G$ and its observer $det(G)$, an edit function $f_e$ is privately safe if $\forall s \in \LANG (det(G))$, $f_e(s) \in L_{\textit{safe}}$.
	\end{definition}

	Recently a three-player game structure called \emph{three-player observer (TPO)} w.r.t.\ the system was defined in~\cite{ji2019synthesis} to embed edit functions. For the sake of completeness, we recall this definition (more details are available in~\cite{ji2019synthesis}).
	\begin{definition}[Three-player Observer] \label{def:tripartite} 
		Given a system $G$ with its observer $det(G)$ and desired observer $det_d(G)$, let $I\subseteq X_{obsd}\times X_{obs}$ be the set of information states. A three-player observer w.r.t.\ $G$ is a tuple of the form $T=(Q_Y, Q_Z, Q_W, \ACT, \ACT^{r}, \Theta, \trans_{yz}, \trans_{zz}, \trans_{zw}, \trans_{wy}, y_0)$, where:
		\begin{itemize}
			\item 
			$Q_Y \subseteq I$ is the set of $Y$ states.
			\item 
			$Q_Z \subseteq I\times \ACT$ is the set of $Z$ states. 
			Let $I(z)$, $E(z)$ denote the information state component and observable event 
			component of a $Z$ state $z$ respectively, so that $z =(I(z), E(z))$. 
			\item 
			$Q_W \subseteq I\times (\ACT \cup \ACT^{r})$ is the set of $W$-states. 
			Let $I(w)$, $A(w)$ denote the information state component and action component 
			of a $W$ state $w$ respectively, so that $w=(I(w), A(w))$.
			\item 
			$\ACT$ is the set of observable events.
			\item 
			$\ACT^{r}$ is the set of event-erasure events.
			\item 
			$\Theta \subseteq \ACT\cup\{\epsilon\} \cup\ACT^r$ is the set of edit decisions at $Z$ states. 
			\begin{enumerate}
				\item\label{im:yz}
				$\trans_{yz}: Q_Y \times \ACT \times Q_Z$ is the transition function from $Y$ states 
				to $Z$ states. For $y=(x_{d}, x_{f})\in Q_Y$, $e_o\in \ACT $, we have: 
				$
				y\trans[ e_o]_{yz}z\Rightarrow [x_{f}\trans[ e_o]_{obs}]\wedge[I(z)=y]\wedge [E(z)=e_o].
				$
				
				\item \label{im:zz}
				$\trans_{zz}: Q_Z \times \Theta \times Q_Z$ is the transition function from $Z$ states to $Z$ states. For $z=((x_{d}, x_{f}), e_o)		\in Q_Z$, $\theta\in \Theta$, we have:
				$
				z\trans[\theta]_{zz}z'\Rightarrow [\theta \in \ACT]\wedge[I(z')=(x'_{d}, x_{f})] 
				\wedge[x_d\trans[\theta]_{det_d}x'_{d}]\wedge [E(z')=e_o]. 
				$
				\item \label{im:zw1}
				$\trans_{zw1}: Q_Z \times \Theta \times Q_W$ is the $\epsilon$ insertion transition function from $Z$ states to $W$ states. For $z=((x_{d}, x_{f}), e_o)\in Q_Z$, $\theta\in \Theta$ we have:
				$
				z\trans[ \theta]_{zw1}w\Rightarrow [\theta=\epsilon]\wedge [I(w)=I(z)]\wedge [A(w)=e_o]
				\wedge[x_{d}\trans[ e_o]_{det_d}]\wedge[x_{f}\trans[ e_o]_{obs}].
				$
				\item \label{im:zw2}
				$\trans_{zw2}: Q_Z \times \Theta \times Q_W$ is the event erasure transition function from $Z$ states to $W$ states. For $z=((x_{d}, x_{f}), e_o)\in Q_Z$, $\theta \in  \Theta$, we have:
				$
				z\trans[\theta]_{zw2}w \Rightarrow [\theta=e_o\rightarrow\epsilon]\wedge [I(w)=I(z)]
				\wedge [A(w)=e_o\rightarrow\epsilon]\wedge[x_{f}\trans[e_o]_{obs}].
				$
				\item \label{im:wy1}
				$\trans_{wy1}:Q_W\times \ACT\times Q_Y$ is the transition function from $W$ states whose action component is in $\ACT$ to $Y$ states. For $w=((x_{d}, x_{f}), e_o)\in  Q_W$, we have:
				$
				w\trans[e_o]_{wy1}y\Rightarrow [y=(x'_{d}, x'_{f})]\wedge[x_{d}\trans[ e_o]_{det_d} x'_d] 
				\wedge[x_{f}\trans[e_o]_{obs}x'_f].
				$
				\item\label{im:wy2}
				$\trans_{wy2}:  Q_W\times \ACT\times Q_Y$ is the  transition function from $W$ states whose action component is in $\ACT^r$ to $Y$ states. For $w=((x_{d}, x_{f}), e_o\rightarrow \epsilon)\in  Q_W$, we have:
				$
				w\trans[ e_o]_{wy2}y\Rightarrow [y=(x_{d}, x'_{f})]\wedge[x_{f}\trans[e_o]_{obs}x'_f].
				$
				
			\end{enumerate}
			\item $y_0=(x_{obsd,0}, x_{obs,0}) \in Q_Y$ is the initial state of $T$, where $x_{obsd,0}$ and $x_{obs,0}$ are initial states of $det_d(G)$ and $det(G)$, respectively.
		\end{itemize}
	\end{definition}

	In general, a three-player observer characterizes a game between a dummy player, the edit function and the environment (system). 
	The state space of a TPO is partitioned as: $Q_Y$ states ($Y$ states) where the dummy player plays; $Q_Z$ states ($Z$ states) where the edit function plays; $Q_W$ states ($W$ states) where the environment plays. A $Y$ state contains the intruder's estimate (left component) as well as the system's true state estimate (right component). A $\rightarrow_{yz}$ transition is defined out of a $Y$ state, indicating that an observable event may occur and thus is received by the edit function. Then the TPO transits to a $Z$ state and the turn of the game is passed to the edit function. Notice that the observable event does not really occur and this dummy player is only introduced to help determine the decisions of edit functions. 
	
	At a $Z$ state, the edit function may choose to insert certain events (including $\epsilon$) or erase its last observed event. If a non-$\epsilon$ event is inserted, a $\rightarrow_{zz}$ transition leads the TPO to another $Z$ state, which means the edit function still has the turn to insert more events until it decides to stop insertion by inserting $\epsilon$ or by erasing the last observed event. There may be multiple transitions defined out of a $Z$ state, i.e., multiple edit decisions;  we write $\Theta(z)$ to denote the set of edit decisions defined at $z \in Q_Z$ in a TPO.

	If the edit function inserts nothing (respectively erases the event it receives from the dummy player), then a $\rightarrow_{zw1}$ (respectively ($\rightarrow_{zw2}$)) transition is defined and the TPO is at a $W$ state. Then the environment plays by letting the observable event executed from its preceding $Y$ state occur. Correspondingly, there are also two types of $\rightarrow_{wy}$ transitions, where $\rightarrow_{wy1}$ indicates that the executed observable event will be observed by the intruder while $\rightarrow_{wy2}$ indicates that the executed observable event will not be observed by the intruder since it has been erased by the edit function.

	When the three players take turns to play, the components of each player's states also get updated. From \Defn~\ref{def:tripartite}, a $\rightarrow_{yz}$ transition does not change the state estimates for the intruder or the system since the player at $Y$ states is dummy and the observable events from $Y$ states do not really occur. With a $\rightarrow_{zz}$ transition, only $x_d$ is updated since $x_d$ is the estimate of the intruder and event insertion only alters the observation of the intruder. For $\rightarrow_{zw}$ transitions, we only require the observable event to be defined at $x_d$ or $x_f$. Finally, a $\rightarrow_{wy1}$ transition updates both $x_d$ and $x_f$ while a $\rightarrow_{wy1}$ transition only updates $x_f$ as the intruder does not observe the erased event. 
	To characterize the information flow in a TPO, the notion of \emph{run} is defined in \cite{ji2019synthesis}.
	
	\begin{definition}[Run]\label{def:run}
		In a three-player observer $T$, a run is defined as:
		$\omega=y_0\xrightarrow{e_{0}}z^1_0\xrightarrow{\theta^1_{0}}z^2_0\xrightarrow{\theta^2_{0}}\cdots\xrightarrow{\theta^{m_0-1}_{0}}z^{m_0}_0\xrightarrow{\theta^{m_0}_{0}}w_0 \xrightarrow{e_0}y_1\xrightarrow{e_{1}}z^1_1\xrightarrow{\theta^1_{1}}z^2_1\xrightarrow{\theta^2_{1}}\cdots z^{m_1}_1\xrightarrow{\theta^{m_1}_1} w_1\xrightarrow{e_1}y_2\cdots\xrightarrow{e_{n}}z^1_n\xrightarrow{\theta^1_{n}}\cdots z^{m_n}_n\xrightarrow{\theta^{m_n}_n}w_n\xrightarrow{e_n}y_{n+1}$, where $y_0$ is the initial state of $T$, $e_i\in \ACT$, $\theta^{j}_i\in \Theta(z^j_i)$, $\forall 0\leq i\leq n$, $1\leq j\leq m_i$ and $n\in \mathbb{N}$, $m_i\in \mathbb N^+$. 
	\end{definition}
	%
	
	We let $\Omega_T$ be the set of all runs in a TPO $T$. For simplicity, similar notations as for automata are defined for three-player observers and thus $T\trans[\omega]x$ denotes the existence of a run in a three-player observer. 
	We also review the concepts of \emph{string generated by a run} and \emph{edit projection} defined in~\cite{ji2019synthesis}. 

	\begin{definition}[String Generated by a Run]\label{def:string-run}	
		Given a run $\omega$ as in Definition~\ref{def:run}, the string generated by $\omega$ is defined as: $l(\omega)=\theta^1_{0}\theta^2_{0}\cdots\theta^{m_0-1}_{0}\theta^{m_0}_{0}e_0\theta^1_{1}\cdots \theta^{m_1}_1e_1\cdots e_{n-1}\theta^1_{n}\cdots\theta^{m_n}_{n}e_n$, where $\forall i\leq n$, $\theta^{m_i}_{i}e_i=\epsilon$ if $\theta^{m_i}_{i}=e_i\rightarrow\epsilon$.
	\end{definition}
	
	\begin{definition}[Edit Projection]\label{def:editproj}
		Given TPO $T$ and  run $\omega_T$ as in \Defn~\ref{def:run}, the edit projection $P_e: \Omega\rightarrow \mathcal L(G)$ is defined such that $P_e(\omega_T)=e_0e_1\cdots e_n$.
	\end{definition}

	In a TPO, $y\in Q_Y$ is a \emph{terminating state} if $\not\exists e_o\in \ACT$, ~\mbox{s.t.} $y\trans[e_o]$. 
	And $w\in Q_W$ is a \emph{deadlocking} state if $\not\exists e_o\in \ACT$, s.t.\ $w\trans[e_o]y$. 
	Also $z\in Q_Z$ is a \emph{deadlocking} state if $\not\exists \theta\in \Theta$, s.t.\ $z\trans[\theta]z'$ or $z \trans[\theta]w$. 
	We call a TPO $T$ \emph{complete}~\cite{ji2019synthesis} if  there are no deadlocking $W$ or $Z$ states in $T$ and $\forall s\in \mathcal L(G)$, $\exists \omega\in \Omega_T$, s.t. $P_e(\omega)=s$.

	\begin{definition}[Edit Function Embedded in TPO]\label{def:tobsembed}
		Given a TPO $T$, a deterministic edit function $f_e$ is embedded in $T$ if $\forall s\in \mathcal L(G)$, $\exists \omega\in \Omega_T$, s.t. $P_e(\omega)=s$ and $l(\omega)=f_e(s)$. 
	\end{definition}
	
	Next, we construct the \emph{largest three-player observer} in the sense that all the other three-player observers are subautomata of it. Such a notion is well defined by considering \emph{all} admissible transitions at \emph{every} state of the TPO, according to the respective conditions in \Defn~\ref{def:tripartite}.

	
	
	

	Edit functions are designed to erase genuine events or insert fictitious ones to mislead the intruder. 
	In theory, it is possible to design an edit function that erases all the events of the system, although this is not desirable. To avoid this situation usually the user provides some constraints on the edit functions. The constraint that is considered in this paper is to limit the number of consecutive erasures.

	\begin{definition}[Edit Constraint]\label{def:feasible-dec}
		The edit constraint, denoted by $\Phi$, requires that the edit function should not make $n+1$ consecutive erasures where $n\in \mathbb{N}$.
	\end{definition}


	Finally, we define the \emph{All Edit Structure (AES)}~\cite{ji2019synthesis} by considering the edit constraint. A synthesis procedure was also presented in~\cite{ji2019synthesis} to construct the AES. Notice that the following definition is slightly different from the AES in the preliminary version of this work~\cite{ji2018efficient} since edit constraints are not considered in~\cite{ji2018efficient}. 
	\begin{definition}[All Edit Structure]\label{def:aes}
		Given system $G$, observer $det(G)$ and desired estimator $det_d(G)$, the All Edit Structure is defined to be the largest complete three-player observer w.r.t.\ $G$, which satisfies the edit constraint.
	\end{definition}
	
	From results in~\cite{ji2019synthesis}, private safety is achievable when the AES is not empty by construction. Hereafter, we assume that the AES is non-empty in the following discussion; if it is empty, then opacity cannot be enforced by the mechanism of edit functions.
	It was also proven in~\cite{ji2019synthesis} that \emph{all} privately safe edit functions satisfying edit constraints are embedded in the AES. Formally speaking, the following result holds. 
	
	\begin{theorem}\label{thm:privateaes} 
		Given a system $G$ and its corresponding AES under edit constraint $\Phi$, 
		an edit function $f_e$ is privately safe and satisfies $\Phi$ if and only if $f_e \in AES$.
	\end{theorem}

	

	We end this section by briefly reviewing the pruning process discussed in~\cite{ji2019synthesis} to construct the AES. 
	The presence of edit constraints may preclude some undesired states from the AES, thus leaving some states  without outgoing transitions, i.e.,``deadlock" $Z$ or $W$ states. 
	Those states reflect the inability of the edit function to issue a valid edit decision (for insertion or erasure) while still maintaining opacity for all possible future behaviors, thus should be removed in the pruning process. 
	Moreover, $Y$ states that have transitions to a deadlock $Z$ state need to be pruned as well, since $Y$-states are the states where the system issues an output event and the edit function is not allowed to prevent their occurrence.

	The construction of the AES may also be interpreted as the calculation of a supervisor where the ``plant'' is the largest three-player observer in terms of subautomaton, including all potentially feasible edit decisions without considering edit constraints.  The $Y$ states are considered as marked states. The events labeling transitions from $Y$ states to $Z$ states and from $W$ states to $Y$ states are considered as uncontrollable, while the events labeling transitions from $Z$ states to $Z$ states and $Z$ states to $W$ states are viewed as controllable. 
	We also define the proper specification by considering edit constraints, deleting states that violate them, and taking the trim of the resulting structure. 
	The goal is to calculate the least restrictive, controllable and nonblocking supervisor based on the plant and this specification. 
	Similar processes of pruning game structures akin to TPOs were discussed in prior work, e.g., \cite{wu2014synthesis, ji2018enforcement, ji2019synthesis}.
	We will leverage this approach in the following discussion, but in the framework of \emph{modular supervisory control}.

	\section{COMPOSITIONAL  ABSTRACTION-BASED METHODOLOGY}\label{sec:generalIdea} 

	\tikzstyle{fancytitle} =[draw=black,thick, fill=gray!20]
	\tikzstyle{block} = [rectangle, draw,  
	text centered, rounded corners, minimum height=2em]
	\tikzstyle{put} = [draw=black,thick, rectangle, draw,  
	text centered,  minimum height=2em]
	\tikzstyle{line} = [draw, -latex']
	\tikzstyle{cloud} = [draw, rounded rectangle ,fill=gray!20, node distance=1.7cm,
	minimum height=2em]
	\tikzstyle{bigbox} = [draw=black,thick, rectangle,minimum width=8.5cm]
	\newcommand\TBox[2][]{%
		\tikz\node[draw,align=left,#1] {#2};\hskip2pt}
	\begin{figure}
		\center
		\includegraphics[scale=0.8]{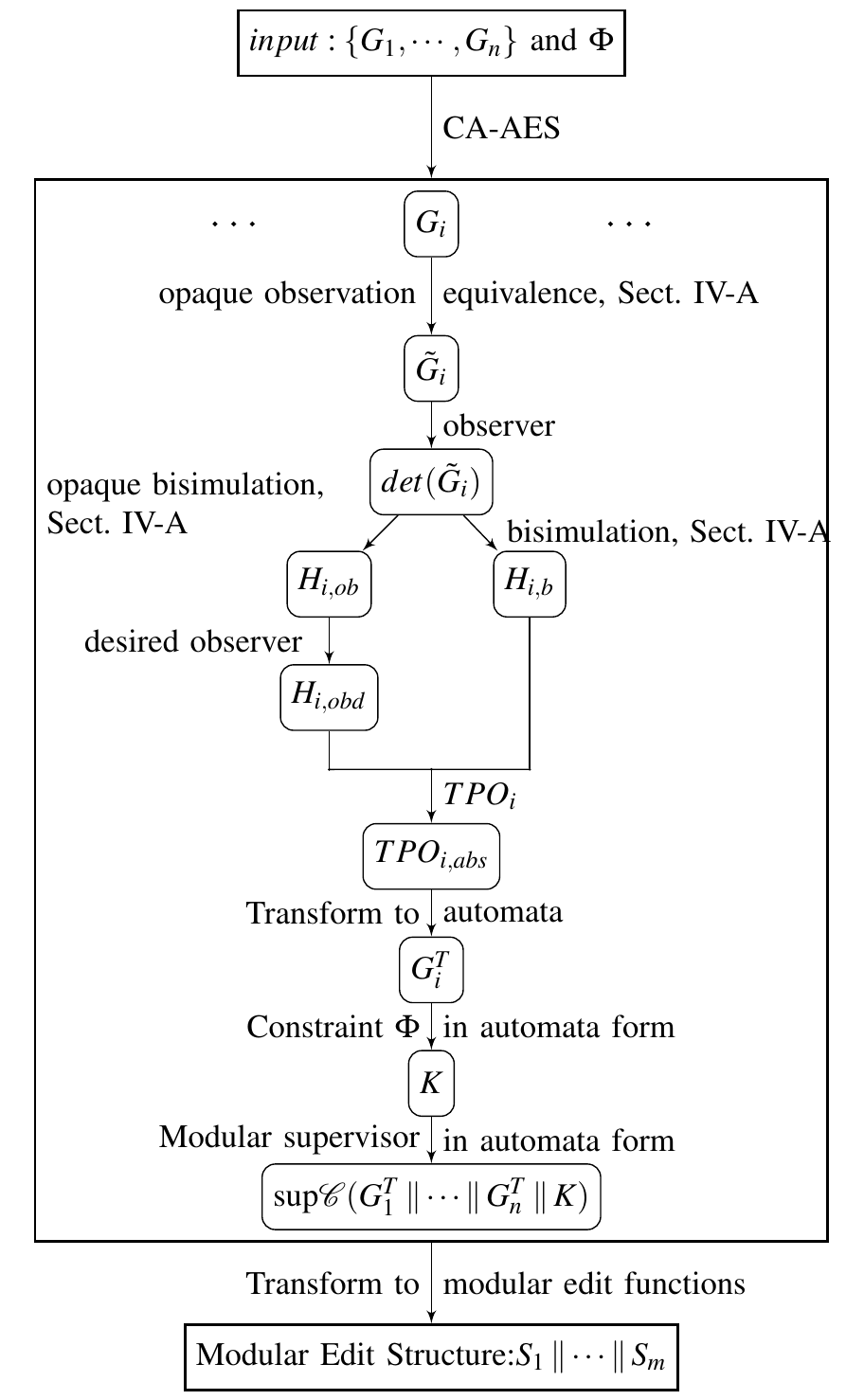}
		\caption{The steps of Algorithm CA-AES.}\label{fig:composAlg}
	\end{figure}
	
	This section presents our novel compositional and abstraction-based methodology for synthesizing modular form edit functions based on individual three-player observers after abstracting the original system.
	For simplicity, we call this methodology the CA-AES (Composition Abstraction-All Edit Structure) Algorithm hereafter.
	The input of the algorithm is a set of nondetermistic automata, $\SYSG=\{G_1, \ldots, G_n\}$ and  
	the output is a modular representation of edit functions, which is called \emph{Modular Edit Structure}.
	The algorithm is summarized in Figure~\ref{fig:composAlg} and its steps are as follows.
	We will explain how to interpret the modular representation of edit functions later.
	
	\begin{enumerate}
		\item
		The algorithm first 
		abstracts each individual automaton, $G_i$, using \emph{opaque observation equivalence}. 
		This results in $\tilde{G}_i$, which has fewer states and transitions compared to the original automaton. 
		\item
		Next, we abstract the observer of $\tilde{G}_i$, i.e., $det(\tilde{G}_i)$, by \emph{opaque bisimulation} and \emph{bisimulation}, resulting in two abstracted deterministic automata $H_{i,ob}$ and $H_{i,b}$. 
		\item
		Then we calculate the abstracted desired observer of $G_i$ from $H_{i,ob}$, which is denoted by $H_{i,obd}$.  
		\item
		Afterward, the largest (abstracted) three-player observer of each individual component $G_i$ is calculated from the abstracted observer $H_{i,b}$ and the abstracted desired observer $H_{i,obd}$, and it is  denoted by $TPO_i$. 

		
		\item
		The final step is to calculate a modular nonblocking and controllable supervisor, then obtain a set of modular edit functions. This is done by transforming the largest three-player observers and the edit constraint to a set of automata, i.e., $G_i^T$ and $K$, respectively. 
		This modular approach is in contrast to calculating monolithic edit functions embedded in the monolithic AES~\cite{ji2019synthesis}. 
	\end{enumerate}
	

	Specifically, in step (iv),  each abstracted three-player observer w.r.t. the corresponding individual system together with the constraint $\Phi$ are transformed to a set of interacting automata. Then in step (v), a modular supremal controllable and nonblocking supervisor is calculated, thereby fulfilling the edit constraint in the composed structure.
	Consequently, the Modular Edit Structure is itself a modular supervisor.
	Regarding step (v), it is possible to leverage existing efficient algorithms on modular supervisory control to calculate the Modular Edit Structure.

	In the monolithic approach of calculating the AES, individual systems $G_1$ through $G_n$ are synchronized first and then the observer of the synchronized system is built.  
	Since the computational complexity of calculating the observer is exponential, synchronizing individual components before building the observer significantly increases the complexity, which may be $2^{\prod_{i=1}^n|Q_i|}$ in the worst case where $|Q_i|$ is the cardinality of $Q_i$. Moreover, constructing the All Edit Structure is polynomial in terms of the state space of the observer, which may be potentially intractable when we deal with the synchronized system. 
	In contrast, our compositional and abstraction-based approach reduces computational cost considerably both from abstracting individual systems and conducting computation in a modular way. 
	However, as will be demonstrated later, some edit decisions may be omitted in the Modular Edit Structure output by Algorithm CA-AES. 
	
	The presented approach relies heavily on the use of three-player observers. We present an example to better understand the structure of such observers.

	\begin{example}
	Consider the nondeterministic automaton $G_1$ with secret states set $Q_1^S=\{q_3\}$, shown in \Fig~\ref{fig:autOFtob}. To generate the three-player observer of $G_1$, first the observer of $G_1$ needs to be built, which is shown as $det(G_1)$ in \Fig~\ref{fig:autOFtob}. To generate the desired observer the state $\{q_3\}\subseteq Q_1^S$ needs to be removed. The desired observer  $det_d(G_1)$ is shown in \Fig~\ref{fig:autOFtob}.  
	
	Then we follow the procedures in~\cite{ji2019synthesis} to build the TPO w.r.t. $det(G_1)$ in \Fig~\ref{fig:autOFtob} (labeled as $T'_1$). As is discussed, the game on the TPO is initiated from $Y$-state $(q_0, q_0)$ where the dummy player executes the observable event $\gamma$ (the only event defined at $q_0$ in $det(G_1)$). Then the edit function takes the turn to play at the $Z$ state $(q_0, q_0, \gamma)$ where it has two choices: insert nothing or erase $\gamma$. If $\gamma$ is erased, then the $W$ state $(q_0, q_0, \gamma\rightarrow \epsilon)$ is reached where the environment plays by executing $\gamma$. Then the turn is passed back to the dummy player and the rest of the structure is interpreted similarly. 

	
	\end{example}

	The compositional abstraction-based approach is explained in more details in the following sections. 
	First, in Section~\ref{sec:AES}, we discuss abstractions at the component level and 
	synchronization of individual three-player observers, formalizing steps (i)-(iv) of Algorithm CA-AES.
	Then, in Section~\ref{sec:sup}, we discuss the last step of Algorithm CA-AES.

	\section{SYNCHRONIZATION AND ABSTRACTION OPERATIONS}\label{sec:AES} 
	
	This section presents results on abstraction and composition that support steps (i)-(iv) of Algorithm CA-CAS.
	First, \Sect~\ref{sec:obsEq} describes the methods to abstract nondeterministic automata and their observers. Next, \Sect~\ref{sec:syncTOB} describes the process of transforming every individual three-player observer to an automaton form and shows that the automaton representation is a substructure (in the sense of subgraph) of the largest monolithic three-player observer.
	
	\subsection{Opaque observation equivalence}\label{sec:obsEq}


	The first strategy used in Algorithm CA-AES to alleviate state space explosion is abstraction of system components.  
	This subsection contains a collection of abstraction methods that can be used to abstract 
	nondeterministc automata and their observers such that the abstracted observers and the desired observers are bisimilar to their original counterparts. 
	The abstraction methods are based on bisimulation and observation equivalence, which are computationally 
	efficient and can be calculated in polynomial-time~\cite{fernandez1990implementation}. 
	We will prove in Theorem~\ref{thm:trisamerun} that if we build the largest three-player observer based on the abstracted observer and the desired observer, we obtain the same runs, consequently the same edit functions as we do from the largest three-player observer based on the original observer and desired observer. 
	
	Bisimulation is a widely-used notion of abstraction that merges states with 
	the same future behavior. 
	\begin{definition}\label{def:bisimgeneral}\cite{milner1989communication}
		Let $G =\langle \Sigma_{\tau}, Q, \rightarrow, Q^0\rangle$ be a nondeterministic automaton.
		An equivalence relation $\approx\ \subseteq Q \times Q$ is called a
		\emph{bisimulation} on~$G$,  if the
		following holds for all $x_1,x_2 \in Q$ such that $x_1 \approx x_2$:
		if  $x_1 \trans[\sigma] y_1$ for some $\sigma \in \ACT_{\tau}$, then there
		exists $y_2 \in Q$  such that $x_2 \trans[\sigma] y_2$, and $y_1\approx y_2$.
	\end{definition}
	
	Bisimulation seeks to merge states with the same outgoing transitions to 
	equivalent states \emph{including unobservable events}, i.e., $\tau$ events. 
	If the unobservable 
	events are disregarded, a more general abstraction 
	method called \emph{weak bisimulation} or \emph{observation equivalence} naturally comes~\cite{milner1989communication}. 
	\begin{definition}\label{def:obseqgeneral}
		Let $G =\langle \Sigma_{\tau}, Q, \rightarrow, Q^0\rangle$ be a nondeterministic automaton.
		An equivalence relation $\sim\ \subseteq Q \times Q$ is called an
		\emph{ observation equivalence} on~$G$,  if the
		following holds for all $x_1,x_2 \in Q$ such that $x_1 \sim x_2$:
		if  $x_1 \ttrans[s] y_1$ for some $s \in \ACTstar$, then there 
		exists $y_2 \in Q$  such that $x_2 \ttrans[s] y_2$, and $y_1\sim y_2$.
	\end{definition}
	
	In order to use observation equivalence for abstraction in the opacity setting, the set of secret states needs to be taken into account. 
	In the following discussion, a restricted version of observation equivalence called \emph{opaque observation equivalence} is employed.
	This notion was first defined in \cite{mohajerani2018transform} in the context of verifying opacity. 
	
	
	\begin{definition}\label{def:obseq}
		Let $G =\langle \Sigma_{\tau}, Q, \rightarrow, Q^0\rangle$ be a nondeterministic automaton 
		with  set of secret states $Q^S\subseteq Q$ and  set 
		of non-secret states $Q^{NS}=Q\setminus Q^S$.
		An equivalence relation $\inobseq_o \subseteq Q \times Q$ is called an
		\emph{opaque observation equivalence} on~$G$ with respects to $Q^S$, if the
		following holds for all $x_1,x_2 \in Q$ such that $x_1\obseq_o x_2$:
		\begin{enumerate}
			\item if
			$x_1 \ttrans[s] y_1$ for some $s \in \ACTstar$, then there
			exists $y_2 \in Q$  such that $x_2 \ttrans[s] y_2$, and $y_1\obseq_o y_2$,
			\item $x_1\in Q^S$ if and only if $x_2\in Q^S$.
		\end{enumerate}  
	\end{definition}
	
	We also wish to use bisimulation to abstract the observer of a nondeterministic 
	system. Besides opaque observation equivalence, \emph{opaque bisimulation} is also defined.

	\begin{definition}\label{def:opaqueBisim}
		Let $G =\langle \Sigma_{\tau}, Q, \rightarrow, Q^0\rangle$ be a nondeterministic 
		automaton with  set of secret states $Q^S\subseteq Q$ 
		and  set of non-secret states $Q^{NS}=Q\setminus Q^S$. 
		Let $det(G)=\langle \ACT, X_{obs}, \trans_{obs},X_{obs}^0\rangle$ be the observer of $G$. 
		An equivalence relation $\approx_{o} \subseteq X_{obs} \times X_{obs}$ is called an
		\emph{opaque bisimulation equivalence} on~$det(G)$ with respects to $Q^S$, if the
		following holds for all $X_1,X_2 \in X_{obs}$ such that $X_1\approx_{o} X_2$:
		\begin{enumerate}
			\item if
			$X_1 \trans[s] Y_1$ for some $s \in \ACTstar$, then there
			exists $Y_2 \in X_{obs}$  such that $X_2 \trans[s] Y_2$, and $Y_1\approx_{o} Y_2$,
			\item $X_1\subseteq Q^S$ if and only if $X_2\subseteq Q^S$.
		\end{enumerate}  
	\end{definition}
	
	The first step of Algorithm CA-AES is to abstract the system using opaque observation equivalence. It has been shown in \cite{rutten1998automata} that if two automata are bisimilar, 
	then their observers are also bisimilar. In this paper this result is extended such that abstracting a nondeterministic automaton using  opaque observation equivalence results in an observer and a desired observer which are bisimilar to the observer and the desired observer of the original system, respectively.

	\begin{proposition}\label{propos:obsBisim}
		Let $G =\langle \Sigma_{\tau}, Q, \rightarrow, Q^0\rangle$ be a nondeterministic automaton with  set of secret states $Q^S\subseteq Q$ and  set of non-secret states $Q^{NS}=Q\setminus Q^S$. Let $\sim_0$ be an opaque observation equivalence on $G$ resulting in $\tilde{G}$ and let $\approx$ be a bisimulation.  Let $det_d(G)$ and $det_d(\tilde{G})$ be the desired observer of $G$ and $\tilde{G}$. Then $det(G)\approx det(\tilde{G})$ and $det_d(G)\approx det_d(\tilde{G})$.
	\end{proposition}
	
	\emph{Proof:} First we prove that $det(G)\approx det(\tilde{G})$. To prove $det(G)\approx det(\tilde{G})$ it is enough to show that $det(G)\trans[s]X$ if and only if $det(\tilde{G})\trans[s]\tilde{X}$, which implies language equivalence between $det(G)$ and $det(\tilde{G})$ since $det(G)$ and $ det(\tilde{G})$ are deterministic. This can be shown by induction. Moreover, in the induction we also show that $x\in X$ if and only if there exist $[x']\in \tilde{X}$ such that $x\in [x']$. This is used for the second part of the proof, where we show  $det_d(G)\approx det_d(\tilde{G})$. 
	
	It is shown by induction on $n\geq 0$ that $X^0\trans[\sigma_1]X^1\trans[\sigma_2]\ldots\trans[\sigma_n]X^n$ in  $det(G)$ if and only if $\tilde{X}^0\trans[\sigma_1]\tilde{X}^1\trans[\sigma_2]\ldots\trans[\sigma_n]\tilde{X}^n$ in $det(\tilde{G})$ such that $x\in X^j$ if and only if $[x']\in \tilde{X}^k$, where $x\in [x']$, for $1\leq j\leq n$.

	\emph{Base case:} $n=0$. Let $X^0$ be the initial state of $det(G)$ and $\tilde{X}^0$ be the initial state of $det(\tilde{G})$. It is shown that $x\in X^0$ if and only if there exists  $[x']\in \tilde{X}^0$ such that $x\in [x']$. 
	
	First, let $x\in X^0$. Then based on $UR(x^0)$, it follows that there exists $x^0\in Q^0$ such that $x^0\ttrans[\tau]x$ in $G$. Since $G\sim_o \tilde{G}$ then based on \Defn~\ref{def:obseq}, there exists $[x'^0]\in\tilde{X}^0$ such that  $[x'^0]\ttrans[\tau] [x']$ in $\tilde{G}$ such that $x^0\in [x'^0]$ and $x\in [x']$. Then based on $UR(x^0)$ it follows that $[x']\in \tilde{X}^0$.

	Now let $[x']\in \tilde{X}^0$. Then based on $UR(x^0)$, it follows that there exists $[x'^0]\in Q^0$ such that $[x'^0]\ttrans[\tau][x']$ in $\tilde{G}$. Since $G\sim_o \tilde{G}$ then based on \Defn~\ref{def:obseq}, there exists $x^0\in X^0$ such that  $x^0\ttrans[\tau] x$ in $G$ such that $x^0\in [x'^0]$ and $x\in [x']$. Then based on $UR(x^0)$ it follows that $x\in X^0$.
	
	\emph{Inductive step}: Assume the claim holds for some $n\geq 0$, i.e,  $X^0\trans[\sigma_1\sigma_2\ldots\sigma_n]X^n=X$ in $det(G)$ if and only if $\tilde{X}=\tilde{X}^0\trans[\sigma_1\sigma_2\ldots\sigma_n]\tilde{X}^n=\tilde{X}$ in $det(\tilde{G})$, such that  $x\in X^k$ if and only if there exists $[x']\in \tilde{X}^k$ such that $x\in [x']$ for all $0\leq k< n$\footnote{Since the base case of the induction is proven for $n=0$, $X^0\trans[\epsilon]$, the inductive step is considered true for $0\leq k< n$.}. It must be shown that $X=X^n\trans[\sigma_{n+1}]Y$ in $det(G)$ if and only if $\tilde{X}=\tilde{X}^n\trans[\sigma_{n+1}]\tilde{Y}$  in $det(\tilde{G})$ such that $x\in X$  if and only if there exists $[x']\in \tilde{X}$ such that $x\in [x']$. 
	
	First, let $X=X^n\trans[\sigma_{n+1}]Y$ in $det(G)$ and let $x\in X$. Then based on $UR(x)$  it holds that $x=x^1\ttrans[\tau]\cdots\ttrans[\tau]x^r\trans[\sigma_{n+1}]y$ in $G$, where $x^j\in X$ for all $1\leq j\leq r$ and $y\in Y$. Since $G\sim_o \tilde{G}$ it holds that $[x']=[x'^1]\ttrans[\tau]\cdots\ttrans[\tau][x'^r]\trans[\sigma_{n+1}][y']$ in $\tilde{G}$ such that $x^j\in [x'^j]$ for all $1\leq j\leq r$ and $y\in [y']$. 
	Based on  $UR(x)$ and inductive assumption it holds that $det(\tilde{G})\trans[\sigma_1\sigma_2\ldots\sigma_n] \tilde{X}^n=\tilde{X}\trans[\sigma_{n+1}]\tilde{Y}$ and $[x']\in \tilde{X}$.
	
	Now let $\tilde{X}=\tilde{X}^n\trans[\sigma_{n+1}]\tilde{Y}$ in $det(\tilde{G})$ and let $[x]\in \tilde{X}$. Then based on $UR(x)$  it holds $[x]=[x^1]\ttrans[\tau]\cdots\ttrans[\tau][x^r]\trans[\sigma_{n+1}][y]$ in $\tilde{G}$, where $[x^i]\in \tilde{X}$ for all $1\leq i\leq r$ and $[y]\in \tilde{Y}$. Since $G\sim_o \tilde{G}$ it holds that $x'=x'^1\ttrans[\tau]\cdots\ttrans[\tau]x'^r\trans[\sigma_{n+1}]y'$ in $\tilde{G}$ such that $x'^i\in [x^i]$ for all $1\leq i\leq r$ and $y'\in [y]$. 
	Based on  $UR(x)$ and inductive assumption it holds that $det(G)\trans[\sigma_1\sigma_2\ldots\sigma_n]X^n=X\trans[\sigma_{n+1}]Y$ such that $x'\in X$.
	
	Now we need to show that $det_d(G)\approx det_d(\tilde{G})$. It was proven above that $det(G)\approx det(\tilde{G})$, which means  $det(G)\trans[s]X$ if and only if $det(\tilde{G})\trans[s]\tilde{X}$ and $x\in X$ if and only if $[x']\in \tilde{X}$, where $x\in [x']$. Therefore, it is enough to show that $X\not\in X_{obsd}$ if and only if $\tilde{X}\not\in \tilde{X}_{obsd}$. 
	
	First assume $X\subseteq Q^S$, which means for all $x\in X$ it holds that $x\in Q^S$ and $X\not\in X_{obsd}$. Since for all $x\in X$ it holds that there exist $[x']\in \tilde{X}$ such that $x\in [x']$ then based on \Defn~\ref{def:obseq} it  holds that $[x']\in \tilde{Q}^S$. Thus, it can be concluded that for all $[x']\in \tilde{X}$ it holds that $[x']\in \tilde{Q}^S$. This means that $\tilde{X}\subseteq \tilde{Q}^S$ and consequently $\tilde{X}\not\in \tilde{X}_{obsd}$.
	
	Now assume $\tilde{X}\subseteq \tilde{Q}^S$, which means for all $[x']\in \tilde{X}$ it holds that $[x']\in \tilde{Q}^S$ and $\tilde{X}\not\in \tilde{X}_{obsd}$. If $[x']\in \tilde{Q}^S$ then for all $x\in [x']$ it holds that $x\in Q^S$. Moreover, it was shown above that $[x']\in \tilde{X}$ if and only if $x\in X$, where $x\in [x']$. Thus, from $\tilde{X}\subseteq \tilde{Q}^S$ it follows that $X\subseteq Q^S$, which means that $X\not\in X_{obsd}$.
	
	Thus, it can be concluded that $det_d(G)\approx det_d(\tilde{G})$.\QEDA
	
	Opaque observation equivalence seeks to merge states of a nondeterministc automaton, which are ``equivalent", before constructing the observer. 
	After calculating the observer, it is possible to further abstract the observer using opaque bisimulation. This guarantees that the smallest abstracted observer generates the same language as the original observer.    
	In the following, Proposition~\ref{pro:bisidetd} shows that if opaque bisimulation is used to abstract the observer, then the abstracted desired observer is also bisimilar to the original desired observer. 
	
	\begin{proposition}
		\label{pro:bisidetd}
		Let $G =\langle \Sigma_{\tau}, Q, \rightarrow, Q^0\rangle$ be a nondeterministic automaton with  set of secret states $Q^S\subseteq Q$ and  set of non-secret states $Q^{NS}=Q\setminus Q^S$. Let $\approx_o$ be an opaque bisimulation  on $det(G)$ resulting in $\widetilde{det(G)}$. Let $det_d(G)$ and $H_{d}$ be the desired observers of $det(G)$ and $\widetilde{det(G)}$, respectively. Then $det_d(G)\approx H_{d}$, where $\approx$ is a bisimulation relation.
	\end{proposition}
	
	\emph{Proof:}
	Since $det(G)\approx_o \widetilde{det(G)}$ based on \Defn~\ref{def:opaqueBisim} it holds that $det(G)\trans[s]X$ if and only if $\widetilde{det(G)}\trans[s][X']$ and $X\in [X']$. Thus, it is enough show that $X\not\in X_{obs, det_d(G)}$ if and only if $[X']\not\in X_{obs,H_d}$, where $X\in [X']$.
	
	First assume $X\subseteq Q^S$, so $X\not\in X_{obs,det_d(G)}$. Then since $X\in [X']$ based on \Defn~\ref{def:opaqueBisim} it holds that for all $X'\in[X']$, $X'\subseteq Q^S$. This means $[X']\subseteq Q^S$ and consequently $[X']\not\in X_{obs,H_d}$.
	
	Then assume $[X']\subseteq Q^S$, so $[X']\not\in X_{obs,H_d}$. Since $X\in [X']$ based on \Defn~\ref{def:opaqueBisim}, $X \subseteq Q^S$ holds, i.e., $X\not\in X_{obs,det_d(G)}$. \QEDA

	We now present the main results of this subsection.
	
	
\def\OpaqueOECS{%
Let $G$ be a nondeterministic automaton with secret states $Q^S\subseteq Q$ and non-secret states $Q^{NS}=Q\setminus Q^S$.  Let $det(G)$ and $det_d(G)$ be the observer and the desired observer of $G$, respectively. Let $\sim_o$ be an opaque observation equivalence on $G$ such that $\tilde{G}\sim_o G$. Let $H_{ob}\approx_o det(\tilde{G})$ and $H_b\approx det(\tilde{G})$ where $\approx_o$ and $\approx$ are opaque bisimulation and bisimulation, respectively. Let $H_{obd}$ be the desired observer of $H_{ob}$. Let $T$ be the largest three-player observer w.r.t.\ $det(G)$ and $det_d(G)$, also let $T'$ be the largest three-player observer w.r.t.\ $H_{obd}$ and $H_b$. Then $T\trans[\omega]q$ if and only if $T'\trans[\omega]\tilde{q}$.}

	\begin{theorem}\label{thm:trisamerun}
	\OpaqueOECS
	\end{theorem}
\emph{Proof:} From Propositions~\ref{propos:obsBisim} and~\ref{pro:bisidetd} it holds that $det(G)\approx H_b$ and $det_d(G)\approx H_{obd}$. Thus, we need to show that  a transition is defined in $T$ if and only if the same transition is defined in $T'$. It is shown by induction on $n\geq 0$ that $y^0\trans[\omega]q_n$ in  $T$ if and only if $\tilde{y}^0\trans[\omega]\tilde{q}_n$ in $T'$.
	
	\emph{Base case:} $(\Rightarrow)$ First assume $y^0\trans[e_o]z^0$ in $T$, where $y^0=(X^0_{obsd},X^0_{obs})$. Based on \Defn~\ref{def:tripartite} it holds that $X^0_{obs}\trans[e_o]$ in $det(G)$, $I(z^0)=y^0$ and $E(z^0)=e_o$. From $X^0_{obs}\trans[e_o]$ in $det(G)$ and since $det(G)\approx \hdet$ it holds that $\tilde{X}^0_{obs}\trans[e_o]$ in $\hdet$. Thus,  $\tilde{y}^0=(\tilde{X}^0_{obsd},\tilde{X}^0_{obs})$ and $\tilde{X}^0_{obs}\trans[e_o]$ in $\hdet$, $I(\tilde{z}^0)=\tilde{y}^0$ and $E(\tilde{z}^0)=e_o$. This means $\tilde{y}^0\trans[e_o]\tilde{z}^0$ in $T'$.
	
	$(\Leftarrow)$ Now assume $\tilde{y}^0\trans[e_o]\tilde{z}^0$ in $T'$, where $\tilde{y}^0=(\tilde{X}^0_{obsd},\tilde{X}^0_{obs})$. The same argument as $(\Rightarrow)$ holds.

	\emph{Inductive step:} Assume the claim holds for some $n\geq 0$, i.e, $n\geq 0$ that $y^0\trans[\omega]q_n$ in  $T$ if and only if $\tilde{y}^0\trans[\omega]\tilde{q}_n$ in $T'$. 
	
	$(\Rightarrow)$ It must be shown that if $q_n\trans[\sigma]p_n$ in $T$ then $\tilde{q}_n\trans[\sigma]\tilde{p}_n$ in $T'$. There are six possibilities:
	\begin{itemize}
		\item $q_n=y$ is a $Y$ state and $p_n=z$ is a $Z$ state, \Defn~\ref{def:tripartite}~\ref{im:yz}. Let  $y=(x_d,x_{f})$ and $\sigma\in \ACT$. Based on inductive assumption there exists $\tilde{y}=(\tilde{x}_{d},\tilde{x}_{f})$ and $\tilde{y}^0\trans[\omega]\tilde{y}$ in $T'$. From $y\trans[\sigma]z$ in $T$ and based on \Defn~\ref{def:tripartite} it holds that $x_{f}\trans[\sigma]$ in $det(G)$ and $I(z)=y$ and $E(z)=\sigma$. Since $det(G)\approx \hdet$ it holds that there exists $\tilde{x}_{f}\trans[\sigma]$ in $\hdet$, where $\tilde{x}_{f}\approx x_f$.  This means $\tilde{y}\trans[\sigma]\tilde{z}$ in $T'$, where $\tilde{y}=(\tilde{x}_{d},\tilde{x}_{f})$ and $I(\tilde{z})=(\tilde{x}_{d},\tilde{x}_{f})$ and $E(\tilde{z})=\sigma$. 
		
		\item $q_n=z$ is a $Z$ state and $p_n=z'$ is a $Z$ state, \Defn~\ref{def:tripartite}~\ref{im:zz}. Let $z=((x_{d},x_{f}),e_o)$ and $\sigma\in\Theta$. Then based on \Defn~\ref{def:tripartite} it holds that $\sigma\in \ACT$ and $I(z')= (x'_{d},x_{f})$ and $x_d\trans[\sigma]x'_d$ in $det_d(G)$ and $E(z')=e_o$.  Since $det_d(G)\approx \hdes$ and from $x_d\trans[\sigma]x'_d$ in $det_d(G)$ it follows  that $\tilde{x}_{d}\trans[\sigma]\tilde{x}'_{d}$ in $\hdes$, where $\tilde{x}_{d}\approx x_d$ and $\tilde{x}'_{d}\approx x'_d$, and based on the inductive assumption it follows that there exists $\tilde{z}=((\tilde{x}_{d},\tilde{x}_{f}), e_o)$ such that $\tilde{y}^0\trans[\omega]\tilde{z}$ in $T'$. Thus,  $\tilde{z}\trans[\sigma]\tilde{z}'$ in $T'$, where $\tilde{z}=((\tilde{x}_{d},\tilde{x}_{f}), e_o)$ and $\sigma\in \ACT$ and $I(\tilde{z}')= (\tilde{x}'_{d},\tilde{x}_{f})$ and $\tilde{x}_d\trans[\sigma]\tilde{x}'_d$ in $\hdes$ and $E(\tilde{z}')=e_o$. 
		
		\item $\tilde{q}_n\trans[\sigma]\tilde{p}_n$ is the $\epsilon$ insertion function and $q_n=z$ is a $Z$ state and $p_n=w$ is a $W$ state, \Defn~\ref{def:tripartite}~\ref{im:zw1}. Let $z=((x_{d},x_{f}),e_o)$ and $\sigma\in\Theta$.  Then based on \Defn~\ref{def:tripartite} it holds that $\sigma=\epsilon$ and $I(w)= I(z)$ and $A(w)=e_o$ and $x_d\trans[e_o]$ and $x_f\trans[e_o]$ in $det_d(G)$ and $det(G)$, respectively. Since $det(G)\approx \hdet $ and $det_d(G)\approx \hdes$, from  $x_f\trans[e_o]$ in $det(G)$ it follows that $\tilde{x}_f\trans[e_o]$ in $\hdet$ and from $x_d\trans[e_o]$ in $det_d(G)$ it follows that   
		$\tilde{x}_d\trans[e_o]$ in $\hdes$. Moreover, based on the inductive assumption it holds that there exists $\tilde{z}=((\tilde{x}_{d},\tilde{x}_{f}), e_o)$ such that $\tilde{y}^0\trans[\omega]\tilde{z}$ in $T'$. 
		Thus, $\tilde{z}\trans[\sigma]\tilde{w}$ in $T'$, where 
		based on \Defn~\ref{def:tripartite} it holds that $\sigma=\epsilon$ and $I(\tilde{w})= I(\tilde{z})$ and $A(\tilde{w})=e_o$ and $\tilde{x}_d\trans[e_o]$ and $\tilde{x}_f\trans[e_o]$ in $\hdes$ and $\hdet$, respectively.
		
		\item $\tilde{q}_n\trans[\sigma]\tilde{p}_n$ is the event erasure transition function and $q_n=z$ is a $Z$ state and $p_n=w$ is a $W$ state, \Defn~\ref{def:tripartite}~\ref{im:zw2}. Let $z=((x_{d},x_{f}),e_o)$ and $\sigma\in\Theta$.  Then based on \Defn~\ref{def:tripartite} it holds that $\sigma=e_o\trans[]\epsilon$ and $I(w)= I(z)$ and $A(w)=e_o\trans[]\epsilon$ and $x_f\trans[e_o]$  in $det(G)$. Since $det(G)\approx \hdet $  from  $x_f\trans[e_o]$ in $det(G)$  it follows that   
		$\tilde{x}_f\trans[e_o]$ in $\hdet$.  Moreover, based on the inductive assumption it follows that there exists $\tilde{z}=((\tilde{x}_{d},\tilde{x}_{f}), e_o)$ such that  $\tilde{y}^0\trans[\omega]\tilde{z}$ in $T'$. 
		Thus, $\tilde{z}\trans[\sigma]\tilde{w}$ in $T'$, where 
		based on \Defn~\ref{def:tripartite} it holds that $\sigma=e_o\trans[]\epsilon$ and $I(\tilde{w})= I(\tilde{z})$ and $A(\tilde{w})=e_o\trans[]\epsilon$ and $\tilde{x}_f\trans[e_o]$ in $\hdet$.
		
		\item  $q_n=w$ is a $W$ state and $p_n=y$ is a $Y$ state and $\sigma\in \ACT$, \Defn~\ref{def:tripartite}~\ref{im:wy1}. Let $w=((x_{d},x_{f}),e_o)$.  Then based on \Defn~\ref{def:tripartite} it holds that $y=(x'_d,x'_f)$ and  $x_f\trans[e_o]x'_f$  in $det(G)$ and $x_d\trans[e_o]x'_d$  in $det_d(G)$. Since $det(G)\approx \hdet$ and $det_d(G)\approx \hdes$, from  $x_f\trans[e_o]x'_f$ in $det(G)$ it follows that $\tilde{x}_f\trans[e_o]\tilde{x}'_f$ in $\hdet$ and from $x_d\trans[e_o]x'_d$ in $det_d(G)$ it follows that   
		$\tilde{x}_d\trans[e_o]\tilde{x}'_d$ in $\hdes$.   Moreover, based on the inductive assumption it holds that there exists $\tilde{w}=((\tilde{x}_{d},\tilde{x}_{f}), e_o)$ such that $\tilde{y}^0\trans[\omega]\tilde{w}$ in $T'$. 
		Thus, $\tilde{w}\trans[\sigma]\tilde{y}$ in $T'$, where 
		based on \Defn~\ref{def:tripartite} it holds that $\tilde{y} = (\tilde{x}'_d,\tilde{x}'_f)$ and $\tilde{x}_f\trans[e_o]\tilde{x}'_f$ in $\hdet$ and $\tilde{x}_d\trans[e_o]\tilde{x}'_d$ in $\hdes$.
		
		\item $q_n=w$ is a $W$ state and $p_n=y$ is a $Y$ state and $\sigma\in \ACT$, \Defn~\ref{def:tripartite}~\ref{im:wy2}. Let $w=((x_{d},x_{f}),e_o\trans[]\epsilon)$.  Then based on \Defn~\ref{def:tripartite} it holds that $y=(x_d,x'_f)$ and  $x_f\trans[e_o]x'_f$  in $det(G)$. Since $det(G)\approx \hdet$ and from  $x_f\trans[e_o]x'_f$ in $det(G)$ it holds that $\tilde{x}_f\trans[e_o]\tilde{x}'_f$ in $\hdet$.  Moreover, based on the inductive assumption it holds that there exists $\tilde{w}=((\tilde{x}_{d},\tilde{x}_{f}), e_o\trans[]\epsilon)$ such that $\tilde{y}^0\trans[\omega]\tilde{w}$ in $T'$. 
		Thus, $\tilde{w}\trans[\sigma]\tilde{y}$ in $T'$, where 
		based on \Defn~\ref{def:tripartite} it holds that $\tilde{y} = (\tilde{x}_d,\tilde{x}'_f)$ and $\tilde{x}_f\trans[e_o]\tilde{x}'_f$ in $\hdet$.
	\end{itemize}
	
	$(\Leftarrow)$ It must be shown that if $\tilde{q}_n\trans[\sigma]\tilde{p}_n$ in $T'$ then $q_n\trans[\sigma]p_n$ in $T$. The same argument as $(\Rightarrow)$ holds. \QEDA
		
		Theorem~\ref{thm:trisamerun} proves that the largest three-player observer obtained from the abstracted system (using opaque observation equivalence and opaque bisimulation) has the same set of runs with that obtained from the original system. This result is essential for the correctness of Algorithm CA-AES. 
	\begin{remark}
	The abstractions in the worst case scenario fail to merge any states. However, as pointed out in the paper the complexity of the abstraction methods is polynomial, while the complexity of calculating the observer is exponential in the number of states. Thus, if the abstraction results in merging even few states, it can potentially reduce the complexity of calculating the observer significantly. Therefore, it is worth applying the abstraction algorithm before calculating the observers.
	\end{remark}
	
	
	\begin{figure}
\includegraphics[width=0.99\columnwidth]{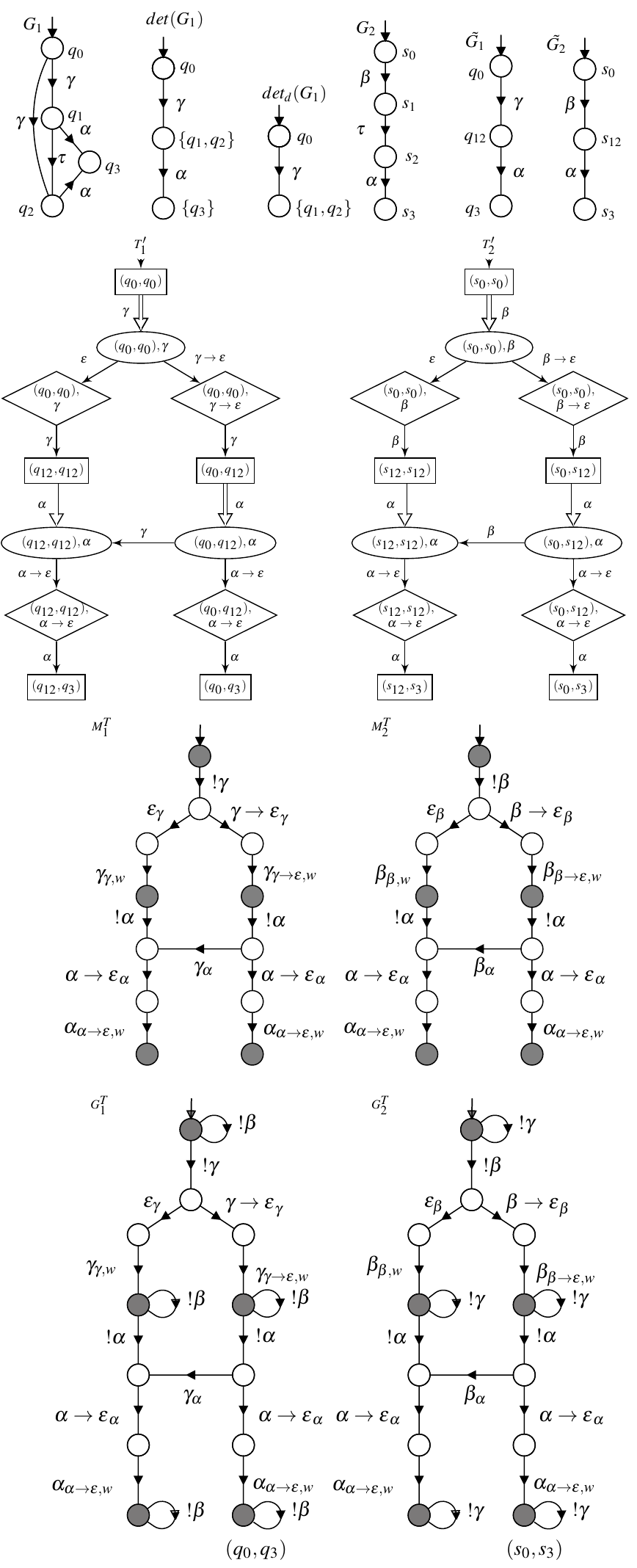}
		\caption{System $\SYSG=\{G_1,G_2\}$ and its abstraction $\{\tilde{G}_1,\tilde{G}_2\}$. The figure also shows the largest three-player observers $T'_1$ and $T'_2$ of the abstracted components and their automata transformations, denoted by $G^T_1$ and $G^T_2$. The uncontrollable events are marked by (!).}\label{fig:autOFtob}
	\end{figure}
	
	\begin{example}\label{ex:abs}
		Consider the nondeterministic system $\SYSG=\{G_1,G_2\}$, shown in \Fig~\ref{fig:autOFtob}, with secret states sets $Q_1^S=\{q_3\}$ and $Q_2^S=\{s_3\}$ where all the events are observable except event $\tau$. In $G_1$ states $q_1$ and $q_2$ are opaque observation equivalent as they both have the same secrecy status and equivalent states can be reached from both, $q_1\trans[\alpha]q_3$ and $q_2\trans[\alpha]q_3$, and $q_1\trans[\tau]q_2$ and $q_2\trans[\epsilon]q_2$. 
		Merging $q_1$ and $q_2$ results in the abstracted automaton $\tilde{G}_1$ shown in \Fig~\ref{fig:autOFtob}. 
		Moreover, states $s_1$ and $s_2$ are also opaque observation equivalent and merging them results in automaton $\tilde{G}_2$ shown \Fig~\ref{fig:autOFtob}. 
		After abstracting the automata, the system becomes a deterministic system. Moreover, the observers as of $\tilde{G}_1$ and $\tilde{G}_2$ are bisimilar to $det(G_1)$ and $det(G_2)$, respectively. The same is also true for the desired observer of $\tilde{G}_1$ and $\tilde{G}_2$. \Fig~\ref{fig:autOFtob} shows the largest three-player observers of $\tilde{G}_1$ and $\tilde{G}_2$, respectively.
		
	\end{example}

	\subsection{Synchronous composition of TPOs}\label{sec:syncTOB} 

	The second strategy used in Algorithm CA-AES to reduce computation complexity is synchronous composition of individual systems. In this work, the main advantage of our compositional approach is to build the largest three-player observer of each component individually, instead of synchronizing individual components and then building the largest monolithic three-player observer. 
	Before synchronization, we first transfer each individual TPO to an automaton using \Defn~\ref{def:TOBtoAutone}. Next,  the individual automata are transformed to a set of interacting automata based on \Defn~\ref{def:TOBtoAut}. 
	It is shown in \Thm~\ref{thm:TOBtoAutSync} that the set of modular three-player observers form a subsystem of their monolithic counterpart, in the sense that some runs are omitted after synchronization. 
	Before \Thm~\ref{thm:TOBtoAutSync}, Lemmas~\ref{propos:syncDet} and~\ref{lem:detd} establish that  synchronization of individual observers (respectively, desired observers) is isomorphic to the observer (respectively, desired observers) of the synchronized system. 
	
	\begin{definition}\label{def:TOBtoAutone}
		Let $T=\langle Q_{Y},Q_{Z},Q_{W},\ACT,\ACT^{\epsilon}, 	\Theta, \trans_{yz},\trans_{zz},\trans_{zw},\trans_{wy},y_0\rangle$ be a three-player observer.
		Automaton $M^T=\langle \ACT_{M^T},Q,\trans,Q^0,Q^m\rangle$ is the monolithic transformed 
		deterministic automaton of $T$ where,
		
		\begin{enumerate}
			\item $\ACT_{M^T}=\ACT\cup[\bigcup\limits_{p\in                 Q_Z}{\Theta}_{E(p)}]\cup[\bigcup\limits_{p\in Q_W }{\ACT }_{A(p),w}] $
			\item $Q =Q_{Y}\cup Q_{Z}\cup Q_{W}$,
			\item $\trans =\{(p,\alpha,q)\mid p\in Q_y\land  p\trans[\alpha]_{yz} q\}\bigcup\{(p,\sigma,q)\mid p\in        Q_Z\land\sigma=\alpha_{E(p)} \land p\trans[\alpha]_{zz, zw_1, zw_2} q\}\bigcup\{(p,\sigma,q)\mid p\in Q_W\land\sigma=\alpha_{A(p),w}\land p\trans[\alpha]_{wy_1, wy_2} q\}$
			\item $Q^0= y_0$,
			\item $Q^m= Q_Y$.
		\end{enumerate}
		The events labeling outgoing transitions mapped from original $Y$ states in $T$, i.e., $\{(p,\alpha,q)\mid p\in Q_y\land  p\trans[\alpha]_{yz} q\}$ and outgoing transitions mapped from original $W$ states in $T$, $\{(p,\sigma,q)\mid p\in Q_W\land\sigma=\alpha_{A(p),w}\land p\trans[\alpha]_{wy_1, wy_2} q\}$ are considered as uncontrollable while the other events are controllable.
	\end{definition}
	
		In \Defn~\ref{def:TOBtoAutone}, $\Sigma_\alpha$ represents that $\alpha$ is added to all the events  of $\ACT$.
	To transform a three-player observer to an automaton, each state of the three-player observer is considered as an automaton state, $Q_i=Q^i_{Y}\cup Q^i_{Z}\cup Q^i_{W}$ in \Defn~\ref{def:TOBtoAutone}.
	Moreover, the information about the states needs to be considered to distinguish some transitions in the transformed automaton and to have a correct synchronization of three-player observers, since the information about state types ($Y$, $W$, $Z$) is lost in the transformation. 
	To this end, in the transformed automaton, the events labeling the transitions from $z$ to $z$ states, from $z$ to $w$ states and from $w$ to $y$ states, need to
have the information of the predecessor states reflected in
them.  Thus, the events in the transformed automaton have the observable event components of $Z$ states, $\bigcup\limits_{p\in                 Q_Z^i}{\Theta^i}_{E(p)}$, and the action components of $W$ states, $\bigcup\limits_{p\in Q_W^i}{\ACT^i}_{A(p),w}$. The initial state of the transformed automaton is the initial state of the three-player observer and the marked  states are the original $Y$ states. 

	\begin{example}\label{ex:TOBtoAutMonolithic}
		Consider the two three-player observers  $T'_1$ and $T'_2$ shown in \Fig~\ref{fig:autOFtob}.  
		To transform the three-player observers  $T_1$ and $T_2$ to their monolithic automata, renaming  $\rho:\{\alpha, \beta, \gamma, \epsilon_\gamma,  {\gamma\to\epsilon}_{\gamma}, \gamma_{\gamma\to\epsilon,w}, \gamma_{\epsilon,w}, \gamma_{\alpha},  \epsilon_\beta, {\beta\to\epsilon}_{\beta}, \beta_{\beta\to\epsilon,w}$, $\beta_{\epsilon,w}, \beta_{\alpha}, {\alpha\to\epsilon}_{\alpha}, \alpha_{\alpha\to\epsilon,w}\}\to \{\alpha, \beta, \gamma, \epsilon, \alpha\to\epsilon,  \beta\to \epsilon, \gamma\to\epsilon\}$ is introduced. Next, 
		all $Y$ states in $T'_1$ and $T'_2$ are considered as marked. \Fig~\ref{fig:autOFtob} shows $M_1^T$ and $M_2^T$, which are the monolithic transformed automata from $T'_1$ and $T'_2$, respectively. All the events that come from transitions defined out of 
		$Y$ states and $W$ states in $T'_1$ and $T'_2$ are considered as uncontrollable in $G_1^T$ and $G_2^T$, thus $\alpha$, $\beta$ and $\gamma$ are uncontrollable. 
	\end{example}

This paper describes the compositional approach for modular systems. In order to have a correct interaction between the transformed automata,  the  transformation of the three-player observers needs to be done in the modular setting.  
	
	\begin{definition}\label{def:TOBtoAut}
		Let $\mathcal{T}=\{T_1,\ldots,T_n\}$ be a three-player observer system such that 
		$T_i=\langle Q^i_{Y},Q^i_{Z},Q^i_{W},\ACT^i,\ACT_i^{\epsilon}, 	\Theta^i, \trans^i_{yz},\trans^i_{zz},\trans^i_{zw},\trans^i_{wy},y^i_0\rangle$. Let $M^T_i=\langle \ACT_{M^T}^i,Q_M^i,\trans_M^i,{Q_M^0}^i,{Q_M^m}^i\rangle$ be the monolithic  transformed deterministic automaton of $T_i$, based on \Defn~\ref{def:TOBtoAutone}. The transformed automaton system of $\mathcal{T}$ is ${\mathcal{G}}^T=\{G_1^T,\cdots,G_n^T\}$, where
		$G^T_i=\langle \ACT_{G^T}^i,Q_i,\trans^i,Q_i^0,Q_i^m\rangle$  and 
		
		\begin{enumerate}
			\item $\ACT_{G^T_i}=\ACT_{M^T}^i\cup[\bigcup\limits_{\alpha\in (\ACT^j\setminus\ACT^i);j\neq i}((\ACT^i\cap \ACT^j)_{\alpha}\cup(\ACT^i\cap \ACT^j)_{\alpha,w}\cup (\ACT^i\cap \ACT^j)_{\alpha\to\epsilon,w})] $
			\item $Q_i=Q_M^i$,
			\item $\trans_M^i\bigcup
			\{(p,\alpha,p)\mid p\in Q_y^i\ \textit{and}\ \alpha\in\bigcup\limits_{j\neq i} (\ACT^j\setminus\ACT^i)\}$
			\item $Q_i^0= {Q_M^0}^i$,
			\item $Q_i^m= {Q_M^m}^i$.
		\end{enumerate}

	\end{definition}

	Since some shared events in the transformed automaton become local after incorporating the extra state information, they need to be added in the alphabet of the transformed automaton, $\bigcup\limits_{\alpha\in (\ACT^j\setminus\ACT^i);j\neq i}((\ACT^i\cap \ACT^j)_{\alpha}\cup(\ACT^i\cap \ACT^j)_{\alpha,w}\cup (\ACT^i\cap \ACT^j)_{\alpha\to\epsilon,w})$ in \Defn~\ref{def:TOBtoAut}. 
	Moreover, the events not defined from $Y$ states of certain TPO $T$ but defined from $Y$ states of some other TPO $T'$ are added as self-loops at the corresponding states in the transformed automaton of $T$, $\{(p,\alpha,p)\mid p\in Q_y^i\ \textit{and}\ \alpha\in\bigcup\limits_{j\neq i} (\ACT^j\setminus\ACT^i)\}$. To create a map between the events of a TPO and its transformed automaton, \emph{renaming} of events is necessary.  Note, when the transformation of a single automaton is considered \Defn~\ref{def:TOBtoAutone} and \Defn~\ref{def:TOBtoAut} produce the same results. Thus, in the following wherever a transformed automaton is discussed we refer to  \Defn~\ref{def:TOBtoAut}. 
	
	Renaming $\rho$ simply removes the extra information from the events of the transformed automaton and maps them back to the original events in the TPO. To be more specific, $\rho$ is a map such that $\rho(\alpha_{\sigma})=\alpha$ and $\rho(\alpha)=\alpha$. Table~\ref{table:events} shows how the events in a transformed automaton are linked to the original events of a TPO, while the third column shows how renaming works. Specifically, in the case where events label $\rightarrow_{yz}$ transitions, renaming does not change events names.

	\begin{table}[tp]
		\def\vph{\vrule width0pt height1.5ex}
		\tabcolsep0.4em
		\extrarowheight0.8pt
		\def\zph{\hphantom{0}}
		\setbox0=\hbox{%
			\begin{tabular}{|>{\sf\bfseries}c|c|c|}
				\hline
				\multicolumn{1}{|c|}{\bf TPO $T$} & \multicolumn{1}{c|}{\bf Automaton $G^T$}& \multicolumn{1}{c|}{\bf Renaming $\rho$}\\
				\hline
				$y\trans[\alpha]_{yz}z$& $y\trans[\alpha]z$ & $\rho(\alpha) = \alpha$\\
				$z\trans[\alpha]_{zz}z'$, $E(z)=e_o$& $z\trans[\alpha_{e_o}]z'$& $\rho(\alpha_{e_o}) = \alpha$ \\
				$z\trans[\alpha]_{zw1}w$, $E(z)=e_o$& $z\trans[\alpha_{e_o}]w$& $\rho(\alpha_{e_o}) = \alpha$\\
				$z\trans[\alpha]_{zw2}w$, $E(z)=e_o$& $z\trans[\alpha_{e_o}]w$& $\rho(\alpha_{e_o}) = \alpha$\\
				$w\trans[\alpha]_{wy1}y$, $A(w)=e_o$& $w\trans[\alpha_{e_o,w}]y$& $\rho(\alpha_{e_o,w}) = \alpha$\\
				$w\trans[\alpha]_{wy2}y$, $A(w)=e_o\to\epsilon$& $w\trans[\alpha_{e_o\to\epsilon,w}]y$& $\rho(\alpha_{e_o\to\epsilon,w}) = \alpha$\\
				\hline
		\end{tabular}}%
		\caption{The table shows the link between the events of a TPO,  its transformed automaton and the corresponding renaming.}%
		\label{table:events}%
		\centerline{\box0}
	\end{table}

	
	\begin{example}\label{ex:TOBtoAut}
		
		Consider the abstracted system $\tilde{\SYSG}=\{\tilde{G}_1,\tilde{G}_2\}$, shown in \Fig~\ref{fig:autOFtob}. The sets of secret states are $\tilde{Q}^S_1=\{q_3\}$, $\tilde{Q}^S_2=\{s_3\}$ where all the events are observable. $T'_1$ and $T'_2$ are the largest three-player observers of $\tilde{G}_1$, $\tilde{G}_2$, respectively.   In Example~\ref{ex:TOBtoAutMonolithic}  the monolithic transformed automata  of $T'_1$ and $T_2'$ were generated. 
		The three-player observer system $\{T'_1, T'_2\}$ is transformed to automata system $\mathcal{G}=\{G^T_1,G^T_2\}$, shown \Fig~\ref{fig:autOFtob}, by adding self-loops at the marked states. Event $\beta$ is not in the alphabet of $T'_1$ so it appears as a self-loop at all marked states in $G^T_1$, which correspond to $Y$ states in $T'_1$. Similarly, $\gamma$ is added as a self-loop at marked states in $G^T_2$ since $\gamma$ is not in the alphabet of $T'_2$. 
	\end{example}

	In the following, Theorem~\ref{thm:TOBtoAutSync} proves that if the synchronization of transformed individual three-player observers contains a transition, then the largest monolithic three-player observer w.r.t.\ the synchronized system also contains an equivalent transition. However, the inverse is not necessarily true as there are some behaviors in the monolithic three-player observer that are omitted in the modular structure. Before Theorem~\ref{thm:TOBtoAutSync}, Lemma~\ref{propos:syncDet}~\cite{SeaSilvaSch:12} and Lemma~\ref{lem:detd} establish that the modular observer and desired observer are isomorphic to their monolithic counterparts. 
	
	\begin{lemma}\label{propos:syncDet}\cite{SeaSilvaSch:12}
		Let $G_1 = \langle\Sigma_1, Q_1, \rightarrow_1, Q^0_1 \rangle$ and $G_2 =\langle\Sigma_2, Q_2, \rightarrow_2, Q^0_2 \rangle$ be two nondeterministic automata. Then $det(G_1\sync G_2)$ is isomorphic to $det(G_1)\sync det(G_2)$.
	\end{lemma}
	
	\begin{lemma}\label{lem:detd}
		Let $G_1 = \langle Q_1, \Sigma_1, \rightarrow_1, Q^0_1\rangle$ and $G_2 = \langle Q_2, \Sigma_2, \rightarrow_2, Q^0_2\rangle$ be two nondeterministic automata with sets of secret states $Q_1^S$ and $Q_2^S$, respectively. 
		Then $det_d(G_1)\sync det_d(G_2)$ is isomorphic to $det_d(G_1\sync G_2)$. 
	\end{lemma}
	\emph{Proof:} 
	From $det(G_1\sync G_2)$ is isomorphic to $det(G_1)\sync det(G_2)$ it follows that $det(G_1\sync G_2)\trans[s]X$ if and only if $det(G_1)\sync det(G_2)\trans[s](X_1,X_2)$ and $(x_1,x_2)\in X$ if and only if $(x_1,x_2)\in X_1\times X_2$. Now we need to show that $X\not\in X_{obsd}$ if and only if $(X_1,X_2)\not\in X_{1,obsd}\times X_{2,obsd}$. 
	
	First assume $X\not\in X_{obsd}$, which means $X\subseteq Q^S$. This further means that for all $(x_1,x_2)\in X$, either $x_1\in Q^S_1$ or $x_2\in Q^S_2$, which implies either $X_1\not\in X_{1,obsd}$ or $X_2\not\in X_{2,obsd}$. Thus, $(X_1,X_2)\not\in X_{1,obsd}\times X_{2,obsd}$.
	
	
	Now assume $(X_1,X_2)\not\in X_{1,obsd}\times X_{2,obsd}$. This means either $X_1\not\in X_{1,obsd}$ or $X_2\not\in X_{2,obsd}$, which implies either $X_1\subset Q^S_1$ or $X_2\subset Q^S_2$. Hence for all $(x_1,x_2)\in (X_1,X_2)=X$ either $x_1\in Q^S_1$ or $x_2\in Q^S_2$, which implies $X\subseteq Q^S$. Thus, $X\not\in X_{obsd}$.\QEDA

	\def\TOBtoAutSync{	Let $G_1 = \langle Q_1, \Sigma_1, \rightarrow_1, Q^0_1\rangle$ and $G_2 = \langle Q_2, \Sigma_2, \rightarrow_2, Q^0_2\rangle$ be two nondeterministic automata with sets of secret states $Q_1^S$ and $Q_2^S$, respectively. Let
		$T_1=\langle Q^1_{Y},Q^1_{Z},Q^1_{W},\ACT_1,\ACT_1^{\epsilon}, 
		\Theta_1, \trans^1_{yz},\trans^1_{zz},\trans^1_{zw},\trans^1_{wy},y^1_0\rangle$
		and $T_2=\langle Q^2_{Y},Q^2_{Z},Q^2_{W},\ACT_2,\ACT_2^{\epsilon}, 
		\Theta_2, \trans^2_{yz},\trans^2_{zz},\trans^2_{zw},\trans^2_{wy},y^2_0\rangle$ 
		be the largest three-player observers w.r.t.\ $G_1$ and $G_2$, respectively. Let  $G^T_1=\langle \ACT_{G^T_1},Q_1, \trans^1,Q_1^0, Q_1^m\rangle$ and 
		$G^T_2=\langle \ACT_{G^T_2},Q_2,\trans^2,Q_2^0, Q_2^m\rangle$ be  
		the transformed automata of $T_1$ and $T_2$, respectively. Let $T$ be the largest monolithic three-player observer w.r.t.\ $G_1||G_2$. Then let $\rho:(\ACT_{G^T_1}\cup\ACT_{G_2^T})\to (\ACT_1\cup \ACT_1^{\epsilon}\cup 
		\Theta_1)\cup (\ACT_2\cup \ACT_2^{\epsilon}\cup \Theta_2)$ be a renaming. 
		We have  $[G^T_1\sync G^T_2\trans[s](q_1,q_2)]\Rightarrow [T\trans[\rho(s)]q]$. }
	\begin{theorem}\label{thm:TOBtoAutSync}
	\TOBtoAutSync
	\end{theorem}	\emph{Proof:} We need to show that a transition is defined in $G^T_1\sync G^T_2$ if  the equivalent transition is defined in $T$. It is shown by induction on $n\geq 0$ that $(y_1^0,y_2^0)\trans[s](q_1^n,q_2^n)$ in $G^T_1\sync G^T_2$ implies $y_T^0\trans[\rho(s)]q_T$ in $T$.
	
	Let $G^T_1\sync G^T_2\trans[s](q_1,q_2)$. 
	
	\emph{Base case:} $n=0$. Let $(y_1^0,y_2^0)$ be the initial state of  $G_1\sync G_2$, i.e., $y_1^0$ (respectively $y^0_2$) is the initial state of $T_1$ (respectively $T_2$). From \Defn~\ref{def:tripartite}, $y_1^0=(X^0_{1,obsd},X^0_{1,obs})$ and $y_2^0=(X^0_{2,obsd},X^0_{2,obs})$, where $X^0_{i,obs}$ and $X^0_{i,obsd}$ are the initial state of $det(G_i)$ and $det_d(G_i)$ for  $i\in \{1,2\}$, respectively. From Lemmas~\ref{propos:syncDet} and~\ref{lem:detd}, $det(G_1\sync G_2)$ and $det_d(G_1\sync G_2)$ are isomorphic to $det(G_1)\sync det(G_2)$ and $det_d(G_1)\sync det_d(G_2)$, which implies  $(X^0_{1,obs},X^0_{2,obs})$ is the initial state of $det(G_1)\sync det(G_2)$ and $det_d(G_1)\sync det_d(G_2)$. Thus $y_T^0=((X^0_{1,obsd},X^0_{2,obsd}),(X^0_{1,obs},X^0_{2,obs}))$ is the initial state of $T$.

	\emph{Inductive step:} Assume the claim holds for some $0\leq n$, which means if $(y_1^0,y_2^0)=(q_1^0,q_2^0)\trans[\sigma_0\ldots\sigma_{n-1}](q_1^n,q_2^n)$ in  $G^T_1\sync G^T_2$ then $y^0\trans[\rho(\sigma_0\ldots\sigma_{n-1})]q^n$ in $T$. Now we need to show that if $(q_1,q_2)=(q_1^n,q_2^n)\trans[\sigma_n](p_1,p_2)$ in $G^T_1\sync G^T_2$ then $q=q^n\trans[\rho(\sigma_n)]p$ in $T$. From  $(q_1,q_2)=(q_1^n,q_2^n)\trans[\sigma_n](p_1,p_2)$ in $G^T_1\sync G^T_2$ and based on \Defn~\ref{def:synch} it holds that $q_i\trans[\sigma_n]p_i$ in $G^T_i$ for $i\in \{1,2\}$, which means $q_i\trans[\sigma_n]p_i$ in $T_i$ for  $i\in \{1,2\}$. Consider the following four cases for all the possible transitions:
	\begin{itemize}
		\item if $\rho(\sigma_n)=\sigma_n$ then  based on \Defn~\ref{def:TOBtoAut} it holds that $q_i\trans[\sigma_n]p_i$ is a $yz$ transition in the original $T_i$ such that $E(p_i)=\sigma_n$ and $I(p_i)=q_i$ if $\sigma_n\in\ACT_i$ and $q_i=p_i$ otherwise for $i\in \{1,2\}$.  Let $q_i=(x_{i,d},x_{i,f})$ for  $i\in \{1,2\}$. Based on \Defn~\ref{def:tripartite} this means $x_{i,f}\trans[\sigma_n]$ in $det(G_i)$ if $\sigma_n\in\ACT_i$. Moreover, based on the inductive assumption there exists $y=(x_{d},x_{f})$ such that  $y_T^0\trans[\omega]y$ in $T$, which implies $det(G_1\sync G_2)\trans[P_e(\omega)]x_f$. Since based on Lemma~\ref{propos:syncDet} $det(G_1\sync G_2)$ and $det(G_1)\sync det(G_2)$ are isomorphic it holds $det(G_1)\sync det(G_2)\trans[P_e(\omega)]$ and string $P_e(\omega)$ reaches $(x_{1,f},x_{2,f})$ in $det(G_1)\sync det(G_2)$. Thus, based on $x_{i,f}\trans[\sigma_n]$ in $det(G_i)$ it can be deduced that $(x_{1,f},x_{2,f})\trans[\sigma_n]$ in $det(G_1\sync G_2)$, it also implies $(x_{1,f},x_{2,f})\trans[\sigma_n]$ in $det(G_1\sync G_2)$ by Lemma~\ref{propos:syncDet}. This means $q_T^n=q_T\trans[\sigma_n]p_T$ is a $yz$ transition in $T$.
		
		\item if $\rho(\sigma_n)= \alpha$ and $\sigma_n=\alpha_{e_o}$.  Then based on 
		\Defn~\ref{def:TOBtoAut} there are 	three possibilities for $q_i\trans[\rho(\sigma_n)]p_i$ in $T_i$: it is a $zz$ transition or a $zw_1$ transition or a $zw_2$ transition and $e_o=E(q_i)$. Now consider the following cases: 
		\begin{itemize}
			\item [1)]	${\alpha}_{e_o}\in \Sigma_{G_1^T}\setminus \Sigma_{G_2^T}$. This means $q_1\trans[{\alpha}_{e_o}]p_1$ in $G_1^T$, which implies   $q_1\trans[{\sigma_n}]p_1$ in $T_1$. From ${\sigma_n}_{e_o}\not\in  \Sigma_{G_2^T}$ and \defn~\ref{def:TOBtoAut} it follows that $\sigma_n\not\in  \Sigma_{2}$, which implies $q_2=p_2$. 
			Consider the following three cases: If $q_1\trans[\sigma_n]p_1$ in $T_1$ is a $zz$ transition, then based on \Defn~\ref{def:tripartite} it holds that $x_{1,d}\trans[\sigma_n]x'_{1,d}$ in $det_d(G_1)$. If $q_1\trans[\sigma_n]p_1$ in $T_1$ is a $zw1$ transition, then based on \Defn~\ref{def:tripartite} it holds that $x_{1,d}\trans[\sigma_n]$ in $det_d(G_1)$ and $x_{1,f}\trans[\sigma_n]$ in $det(G_1)$.  If $q_1\trans[\sigma_n]p_1$ in $T_1$ is a $zw2$ transition, then based on \Defn~\ref{def:tripartite} it holds that $x_{1,f}\trans[\sigma_n]$ in $det(G_1)$. In all the three cases based on Lemmas~\ref{propos:syncDet} and~\ref{lem:detd}, where they show $det_d(G_1\sync G_2)$ and $ det_d(G_1)\sync det_d(G_2)$ are isomorphic, it holds that   $(x_{1,d},x_{2,d})\trans[\sigma_n](x'_{1,d},x_{2,d})$ in $det_d(G_1\sync G_2)$ and $(x_{1,f},x_{2,f})\trans[\sigma_n]$ in $det(G_1\sync G_2)$. These  mean $q^n=q\trans[\sigma_n]p$ in $T$ is a $zz$ transition if $q_1\trans[\sigma_n]p_1$ in $T_1$ is a $zz$ transition, $q^n=q\trans[\sigma_n]p$ is a $zw1$ transition in $T$ if $q_1\trans[\sigma_n]p_1$ is a $zw1$ transition in $T_1$ and $q^n_T=q_T\trans[\sigma_n]p_T$ is a $zw2$ transition in $T$ if $q_1\trans[\sigma_n]p_1$ is a $zw2$ transition in $T_1$.
			
			\item[2)] ${\sigma_n}_{e_o}\in \Sigma_{G_1^T}\cap \Sigma_{G_2^T}$. This means $q_i\trans[{\sigma_n}_{e_o}]p_i$ in $G_i^T$ for $i\in \{1,2\}$, which implies   $q_i\trans[{\sigma_n}]p_i$ in $T_i$ for  $i\in \{1,2\}$.  Again there are three cases: If $q_i\trans[\sigma_n]p_i$ in $T_i$ for $i=1,2$ is a $zz$ transition. Then based on \Defn~\ref{def:tripartite} it holds that $x_{i,d}\trans[\sigma_n]x'_{i,d}$ in $det_d(G_i)$ for $i=1,2$.  If $q_i\trans[\sigma_n]p_i$ in $T_i$ is a $zw1$ transition for  $i\in \{1,2\}$. Then based on \Defn~\ref{def:tripartite} it holds that $x_{i,d}\trans[\sigma_n]$ in $det_d(G_i)$ and $x_{i,f}\trans[\sigma_n]$ in $det(G_i)$ for $i=1,2$.  If $q_i\trans[\sigma_n]p_i$ in $T_i$ is a $zw2$ transition for $i=1,2$. Then based on \Defn~\ref{def:tripartite} it holds that  $x_{i,f}\trans[\sigma_n]$ in $det(G_i)$ for $i=1,2$. In all the three cases based on Lemmas~\ref{propos:syncDet} and~\ref{lem:detd} where they show $det_d(G_1\sync G_2)$ and $ det_d(G_1)\sync det_d(G_2)$ are isomorphic, it holds that   $(x_{1,d},x_{2,d})\trans[\sigma_n](x'_{1,d},x'_{2,d})$ in $det_d(G_1\sync G_2)$ and $(x_{1,f},x_{2,f})\trans[\sigma_n]$ in $det(G_1\sync G_2)$. These mean $q^n_T=q_T\trans[\sigma_n]p_T$ in $T$ is a $zz$ transition if $q_i\trans[\sigma_n]p_i$ in $T_i$ is a $zz$ transition, $q^n_T=q_T\trans[\sigma_n]p_T$ is a $zw1$ transition in $T$ if $q_i\trans[\sigma_n]p_i$ is a $zw1$ transition in $T_i$ and $q^n_T=q_T\trans[\sigma_n]p_T$ is a $zw2$ transition in $T$ if $q_i\trans[\sigma_n]p_i$ is a $zw2$ transition in $T_i$ for  $i\in \{1,2\}$.
			\item[3)] ${\sigma_n}_{e_o}\in \Sigma_{G_2^T}\setminus \Sigma_{G_1^T}$. The same argument as case 1. 
		\end{itemize}

		\item if $\rho^{-1}(\sigma_n)= {\sigma_n}_{e_o,w}$ then based on 
		\Defn~\ref{def:TOBtoAut} there are again three cases: 
		\begin{itemize}
			\item[1)] ${\sigma_n}_{e_o,w}\in \ACT_{G^T_1}\setminus\ACT_{G^T_2}$. This means 
			$q_1\trans[{\sigma_n}_{e_o,w}]p_1$ in $G^T_1$. From ${\sigma_n}_{e_o,w}\not\in  \Sigma_{G_2^T}$ and \defn~\ref{def:TOBtoAut} it follows that $\sigma_n\not\in  \Sigma_{2}$, which implies $q_2=p_2$. From $\rho^{-1}(\sigma_n)= {\sigma_n}_{e_o,w}$ 
			it holds that $p_1\trans[\sigma_n]q_1$ in $T_1$ is a $wy1$ transition. 
			This means $x_{1,d}\trans[\sigma_n]x'_{1,d}$ in $det_d(G_1)$ and $x_{1,f}\trans[\sigma_n]x'_{1,f}$ in $det(G_1)$. Based on Lemmas~\ref{propos:syncDet} and~\ref{lem:detd}, where they show $det_d(G_1\sync G_2)$ and $ det_d(G_1)\sync det_d(G_2)$ are isomorphic, it holds that   $(x_{1,d},x_{2,d})\trans[\sigma_n](x'_{1,d},x_{2,d})$ in $det_d(G_1\sync G_2)$ and $(x_{1,f},x_{2,f})\trans[\sigma_n](x'_{1,f},x_{2,f})$ in $det(G_1\sync G_2)$. This means $q^n_T=q_T\trans[\sigma_n]p_T$ is a $wy1$ transition in $T$. 
			
			\item[2)] ${\sigma_n}_{e_o,w}
			\in \ACT_{G^T_1}\cap\ACT_{G^T_2}$. Then it follows that 
			$q_i\trans[{\sigma_n}_{e_o,w}]p_i$ in $G^T_i$ for  $i\in \{1,2\}$. 
			This means $q_i\trans[\sigma]p_i$ in $T_i$ is a $wy1$ transition, which implies
			$x_{i,d}\trans[\sigma_n]x'_{i,d}$ in $det_d(G_i)$ and $x_{i,f}\trans[\sigma_n]x'_{i,f}$ in $det(G_i)$ for  $i\in \{1,2\}$. Based on Lemmas~\ref{propos:syncDet} and~\ref{lem:detd}, where they show $det_d(G_1\sync G_2)$ and $ det_d(G_1)\sync det_d(G_2)$ are isomorphic, it holds that   $(x_{1,d},x_{2,d})\trans[\sigma_n](x'_{1,d},x'_{2,d})$ in $det_d(G_1\sync G_2)$ and $(x_{1,f},x_{2,f})\trans[\sigma_n](x'_{1,f},x'_{2,f})$ in $det(G_1\sync G_2)$. This means $q^n_T=q_T\trans[\sigma_n]p_T$ is a $wy1$ transition in $T$.
			
			\item[3)] ${\sigma_n}_{e_o,w}\in \ACT_{G^T_2}\setminus\ACT_{G^T_1}$. The same argument as case 1.
		\end{itemize}

		\item if $\rho^{-1}(\sigma_n)= {\sigma_n}_{e_o\to\epsilon,w}$ then the same argument 
		as above, $\rho^{-1}(\sigma_n)= {\sigma_n}_{e_o,w}$, holds.  \QEDA
	\end{itemize}

	\Thm~\ref{thm:TOBtoAutSync} shows that synchronization of individual transformed three-player observers is a subsystem of the largest monolithic three-player observer. Specifically, if there is a string $s$ in $G^T_1\sync G^T_2$, then there always exists a corresponding path $\rho(s)$ in $T$. 
	The synchronized automaton form TPOs may not always be equal to the monolithic TPO since some $\rightarrow_{zz}$ transitions may not appear in the synchronized system. 
	This happens when a state in the synchronization of TPOs is a combination of an original $Z$ and an original $Y$ state from individual TPOs, while the observable event component of the original $Z$ state is a local event. 
	However, as there is no difference  between local and shared events in the monolithic approach of obtaining TPOs, the largest monolithic TPO contains all possible transitions of edit decisions.  
	The proof of \Thm~\ref{thm:TOBtoAutSync} also illustrates that for every state in the largest monolithic TPO, there exists a corresponding state in the synchronized individual TPOs in automaton form.

	\begin{figure*}
		\begin{scriptsize}
			\psfrag{s0}{$s_0$}
			\psfrag{s4}{$s_4$}
			\psfrag{s1}{$s_1$}
			\psfrag{s12}{$s_{12}$}
			\psfrag{q12}{$q_{12}$}
			\psfrag{s2}{$s_2$}
			\psfrag{s3}{$s_3$}
			\psfrag{q0}{$q_0$}
			\psfrag{q4}{$q_4$}
			\psfrag{q1}{$q_1$}
			\psfrag{q2}{$q_2$}
			\psfrag{q3}{$q_3$}
			\psfrag{a}{$\alpha$}
			\psfrag{b}{$\beta$}
			\psfrag{d}{$\gamma$}
			\psfrag{tau}{$\tau$}
			\psfrag{G1}{$G_1$}
			\psfrag{ag1}{$\tilde{G}_1$}
			\psfrag{ag2}{$\tilde{G}_2$}
			\psfrag{G2}{$G_2$}
			\qquad\qquad
		\end{scriptsize}
		\centering
		\begin{tiny}
			
			\begin{tikzpicture}[node distance = 2cm, auto]
			
			\node [ini] (ini){$T$};
			\node [inirec, below of=ini] (yAA) {$(A,A)$};
			\node [inov,  right= 2.3cm of yAA] (zAAc) {$(A,A),\gamma$};
			\node [dim, below right=1.3cm of zAAc] (wAAec) {$(A,A),$ $\gamma\to\epsilon$};
			\node [dim, below left=1.3cm of zAAc] (wAAe) {$(A,A),$ $\gamma$};
			\node [rec, below of=wAAec] (yAC) {$(A,C)$};		
			\node [rec, below of=wAAe] (yCC) {$(C,C)$};		
			\node [ov, below of=yCC] (zCCb) {$(C,C),\beta$};	
			\node [ov, below of=yAC] (zACb) {$(A,C),\beta$};
			\node [dim, below left=0.4cm and 0.08cm of  zCCb] (wCCe) {$(C,C),$ $\beta$};
			\node [dim, below right=0.4cm and 0.08cm of zCCb] (wCCbe) {$(C,C),$ $\beta\to\epsilon$};
			\node [dim, below left   of= zACb] (wACe) {$(A,C),$ $\beta$};
			\node [dimdo, below right  of= zACb] (wACbe) {$(A,C),$ $\beta\to\epsilon$};

			\node [inov,  left=2.3cm of yAA] (zAAb) {$(A,A),\beta$};
			\node [dim, below right=1.2cm of zAAb] (wAAeb) {$(A,A),$ $\beta\to\epsilon$};
			\node [dim, below left=1.2cm of zAAb] (wAAe2) {$(A,A),$ $\beta$};
			\node [rec, below of=wAAeb] (yAB) {$(A,B)$};		
			\node [rec, below of=wAAe2] (yBB) {$(B,B)$};		
			\node [ov, below of=yBB] (zBBc) {$(B,B),\gamma$};	
			\node [ov, below of=yAB] (zABc) {$(A,B),\gamma$};
			\node [dim, below left  of=zBBc] (wBBe) {$(B,B),$ $\gamma$};
			\node [dim, below right of=zBBc] (wBBce) {$(B,B),$ $\gamma\to\epsilon$};
			\node [dim, below left=0.4cm and 0.08cm  of zABc] (wABe) {$(A,B),$ $\gamma$};
			\node [dimdo, below right=0.4cm and 0.08cm of zABc] (wABce) {$(A,B),$ $\gamma\to\epsilon$};
			
			\node [rec, below=1.5cm of wABe] (yDD) {$(D,D)$};		
			\node [rec, below=1.5cm of wABce] (yBD) {$(B,D)$};		
			\node [rec, below=1.5cm of wCCe] (yCD) {$(C,D)$};		
			\node [recd, below=1.5cm of wCCbe] (yAD) {$(A,D)$};	
			
			\node [ov, below of=yDD] (zDDa) {$(D,D),\alpha$};	
			\node [ov, below of=yBD] (zBDa) {$(B,D),\alpha$};	
			\node [ov, below of=yCD] (zCDa) {$(C,D),\alpha$};	
			\node [ovd, below of=yAD] (zADa) {$(A,D),\alpha$};
			
			\node [dim, below   of=zDDa] (wDDae) {$(D,D),$ $\alpha\to\epsilon$};
			\node [dimdo, below  of=zBDa] (wBDae) {$(B,D),$ $\alpha\to\epsilon$};
			\node [dim, below   of=zCDa] (wCDae) {$(C,D),$ $\alpha\to\epsilon$};
			\node [dimd, below  of=zADa] (wADae) {$(A,D),$ $\alpha\to\epsilon$};
			
			\node [rec, below of=wDDae] (yDE) {$(D,E)$};		
			\node [recd, below of=wBDae] (yBE) {$(B,E)$};		
			\node [rec, below of=wCDae] (yCE) {$(C,E)$};		
			\node [recd, below of=wADae] (yAE) {$(A,E)$};	
			
			\path[line](ini) to node[above]{}(yAA);
			\draw [vecArrow, very thick] (yAA) to node[above,xshift=-3pt]{$\gamma$}(zAAc);
			\draw [innerWhite] (yAA) to (zAAc);		
			\draw [vecArrow] (yAA) to node[above]{}(zAAc);
			\draw [innerWhite] (yAA) to (zAAc);		
			\draw [vecArrow] (yAA) to node[above,xshift=3pt]{$\beta$}(zAAb);
			\draw [innerWhite] (yAA) to (zAAb);		
			\draw [vecArrow] (yAA) to node[above]{}(zAAb);
			\draw [innerWhite] (yAA) to (zAAb);
			\path[line, very thick](zAAc) to node[right,xshift=3pt,yshift=5pt]{$\gamma\to\epsilon$}(wAAec);
			\path[line](zAAc) to node[left,xshift=-3pt,yshift=5pt]{$\epsilon$}(wAAe);
			\path[line](zAAb) to node[right,xshift=3pt,yshift=5pt]{$\beta\to\epsilon$}(wAAeb);
			\path[line](zAAb) to node[left,xshift=-3pt,yshift=5pt]{$\epsilon$}(wAAe2);
			\path[line, very thick](wAAec) to node[right]{$\gamma$}(yAC);
			\path[line](wAAe) to node[left]{$\gamma$}(yCC);
			\path[line](wAAeb) to node[right]{$\beta$}(yAB);
			\path[line](wAAe2) to node[left]{$\beta$}(yBB);
			
			\draw [vecArrow] (yBB) to node[left,xshift=-3pt]{$\gamma$}(zBBc);
			\draw [innerWhite] (yBB) to (zBBc);		
			\draw [vecArrow] (yBB) to node[left]{}(zBBc);
			\draw [innerWhite] (yBB) to (zBBc);	
			\draw [vecArrow] (yAB) to node[right,xshift=3pt]{$\gamma$}(zABc);
			\draw [innerWhite] (yAB) to (zABc);		
			\draw [vecArrow] (yAB) to node[left]{}(zABc);
			\draw [innerWhite] (yAB) to (zABc);	
			\draw [vecArrow] (yCC) to node[left,xshift=-3pt]{$\beta$}(zCCb);
			\draw [innerWhite] (yCC) to (zCCb);		
			\draw [vecArrow] (yCC) to node[left]{}(zCCb);
			\draw [innerWhite] (yCC) to (zCCb);
			\draw [vecArrow, very thick] (yAC) to node[right,xshift=3pt]{$\beta$}(zACb);
			\draw [innerWhite, very thick] (yAC) to (zACb);		
			\draw [vecArrow] (yAC) to node[left]{}(zACb);
			\draw [innerWhite] (yAC) to (zACb);
			
			\path[line](zABc) to node[above]{$\beta$}(zBBc);
			\path[line](zACb) to node[above]{$\gamma$}(zCCb);
			
			\path[line](zBBc) to node[left,xshift=-3pt,yshift=3pt]{$\epsilon$}(wBBe);
			\path[line](zBBc) to node[right,xshift=3pt,yshift=3pt]{$\gamma\to\epsilon$}(wBBce);		
			\path[line](zABc) to node[left,xshift=-3pt,yshift=3pt]{$\epsilon$}(wABe);
			\path[line](zABc) to node[right,xshift=3pt,yshift=3pt]{$\gamma\to\epsilon$}(wABce);		
			\path[line](zCCb) to node[left,xshift=-3pt,yshift=3pt]{$\epsilon$}(wCCe);
			\path[line](zCCb) to node[right,xshift=3pt,yshift=3pt]{$\beta\to\epsilon$}(wCCbe);		
			\path[line, very thick](zACb) to node[left,xshift=-3pt,yshift=3pt]{$\epsilon$}(wACe);
			\path[line](zACb) to node[right,xshift=3pt,yshift=3pt]{$\beta\to\epsilon$}(wACbe);
			
			\path[line](wBBe) to node[left,xshift=-29pt,yshift=14pt]{$\gamma$}(yDD);
			\path[line](wBBce) to node[left,xshift=-29pt,yshift=14pt]{$\gamma$}(yBD);
			\path[line](wABe) to node[left,xshift=-30pt,yshift=15pt]{$\gamma$}(yCD);
			\draw[dashed,->](wABce) to node[left,xshift=-31pt,yshift=19pt]{$\gamma$}(yAD);
			\path[line](wCCe) to node[right,xshift=29pt,yshift=17pt]{$\beta$}(yDD);
			\path[line](wCCbe) to node[right,xshift=-1pt,yshift=12pt]{$\beta$}(yCD);
			\path[line,very thick](wACe) to node[right,xshift=40pt,yshift=13pt]{$\beta$}(yBD);
			\draw[dashed,->](wACbe) to node[right,xshift=20pt,yshift=10pt]{$\beta$}(yAD);

			\draw [vecArrow] (yDD) to node[left,xshift=-3pt, yshift=4pt]{$\alpha$}(zDDa);
			\draw [innerWhite] (yDD) to (zDDa);		
			\draw [vecArrow] (yDD) to node[left]{}(zDDa);
			\draw [innerWhite] (yDD) to (zDDa);	
			\draw [vecArrow] (yBD) to node[left,xshift=-3pt, yshift=4pt]{$\alpha$}(zBDa);
			\draw [innerWhite, very thick] (yBD) to (zBDa);		
			\draw [vecArrow, very thick] (yBD) to node[left]{}(zBDa);
			\draw [innerWhite] (yBD) to (zBDa);	
			\draw [vecArrow] (yCD) to node[right,xshift=3pt, yshift=4pt]{$\alpha$}(zCDa);
			\draw [innerWhite] (yCD) to (zCDa);		
			\draw [vecArrow] (yCD) to node[right]{}(zCDa);
			\draw [innerWhite] (yCD) to (zCDa);
			\draw [dashed, ->] (yAD) to node[right,xshift=3pt, yshift=4pt]{$\alpha$}(zADa);

			\path[line, very thick](zDDa) to node[left,xshift=-3pt]{$\alpha\to\epsilon$}(wDDae);
			\path[line](zBDa) to node[left,xshift=-3pt]{$\alpha\to\epsilon$}(wBDae);
			\path[line](zCDa) to node[right,xshift=3pt]{$\alpha\to\epsilon$}(wCDae);
			\draw[dashed,->](zADa) to node[right,xshift=3pt]{$\alpha\to\epsilon$}(wADae);
			
			\path[line, very thick](wDDae) to node[left,xshift=-3pt]{$\alpha$}(yDE);
			\draw[dashed,->](wBDae) to node[left,xshift=-3pt]{$\alpha$}(yBE);
			\path[line](wCDae) to node[right,xshift=3pt]{$\alpha$}(yCE);
			\draw[dashed,->](wADae) to node[right,xshift=3pt]{$\alpha$}(yAE);
			
			\draw[dashed,->, out=135,in=15](zADa) to node[left,xshift=-3pt]{$\gamma$}(zCDa);
			\draw[dashed,->, out=135,in=15](zADa) to node[left,xshift=-27pt,yshift=-6pt]{$\beta$}(zBDa);
			\path[line, very thick, out=135,in=15](zBDa) to node[right,xshift=-3pt, yshift=-4pt]{$\gamma$}(zDDa);
			\path[line, out=135,in=15](zCDa) to node[right,xshift=5pt, yshift=4pt]{$\beta$}(zDDa);
			\end{tikzpicture}
		\end{tiny}
		\caption{The monolithic largest three-player observer w.r.t.\ to ${G_1}\sync {G_2}$ (also same as that w.r.t.\ $\tilde{G_1}\sync \tilde{G_2}$ in this particular case) in Example~\ref{ex:syncTOB}.}
		\label{fig:syncTOB}
	\end{figure*}
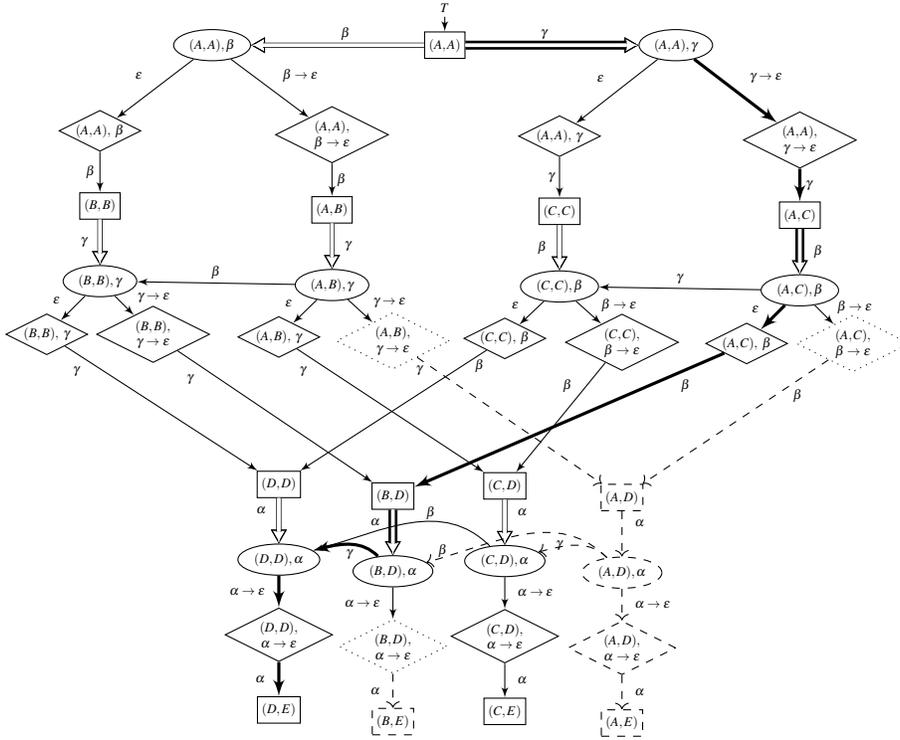

	\begin{remark}
		Although the statement of Theorem~\ref{thm:TOBtoAutSync} concentrates on the case of two individual systems, the result can be generalized to more than two individual systems.  \QEDA
	\end{remark}

	
	\begin{remark}
		Notice that Theorem~\ref{thm:TOBtoAutSync} illustrates that some transitions are ``missing'' in the synchronized automaton compared with the largest monolithic three-player observer $T$. It further implies that more transitions will be missing if we synchronize more individual TPOs (in automaton form).
		Actually, we may locate those missing transitions and add them back to the synchronized automaton $\sync^{n}_{i=1}G^T_i$.
		
		Specifically, consider states $(q_1, q_2, \cdots, q_n)$ and $(q'_1, q'_2, \cdots, q'_n)$ in $\sync^{n}_{i=1}G^T_i$ such that $q_i, q'_i\in Q^i_Y\cup Q^i_Z$ for all $i$, i.e., every component in those states is either a $Y$ state or a $Z$ state from an individual transformed automaton. Then we add transition $(q_1, q_2, \cdots, q_n)\xrightarrow{\sigma}(q'_1, q'_2, \cdots, q'_n)$ if there exists a set of indexes $\mathbb I\in 2^{\{1, 2, \cdots n\}}$, such that for all $i\in \mathbb I$, $q_i=(x_{i,d}, x_{i, f}), q'_i=(x'_{i, d}, x_{i,f})\in Q^i_Y$ and $x_{i,d}\xrightarrow{\sigma}x'_{i,d}$ in $det_d(G_i)$; while for all $i\notin \mathbb I$, $q_i=q'_i$.  Intuitively, the added transition implies that event $\sigma$ may be inserted in the largest monolithic TPO w.r.t.\ $\sync^{n}_{i=1}G_i$. However, due to the fact that there are no transitions defined from a $Y$ state to another $Y$ state in TPOs, those transitions are missing in $\sync^{n}_{i=1}G^T_i$, which implies the synchronized system $\sync^{n}_{i=1}G^T_i$ may only contain a subset of edit decisions in the largest monolithic TPO. 
		
		However, the above mentioned operation may not be preferred in practice since it involves explicitly synchronizing individual TPOs in their automaton form. This is usually not feasible in modular approaches and should be avoided in our Algorithm CA-AES as well.	\QEDA
	\end{remark}


	Finally, the results of this section are formally recapped in Theorem~\ref{pro:trisamerunAfterAbs}, which illustrates that the synchronization of transformed (automaton form) three-player observers w.r.t.\ individual abstracted systems contain a subset of the transitions of the largest monolithic three-player observer. The proof follows directly from Theorem~\ref{thm:TOBtoAutSync} and Theorem~\ref{thm:trisamerun}. 
	
	\begin{theorem}\label{pro:trisamerunAfterAbs}
		Let $G_1$ and $G_2$ be two nondeterministic automaton with sets of secret states $Q_i^S\subseteq Q_i$ and  sets of non-secret states $Q_i^{NS}=Q\setminus Q_i^S$ for $i=1,2$. Let $det(G_i)$ and $det_d(G_i)$ be the observer and the desired observer of $G_i$,  respectively. Let $\sim_o$ be an opaque observation equivalence on $G_i$ 
		such that $\tilde{G}_i\sim_o G_i$ for $i=1,2$. Let $H_{i,ob}\approx_o det(\tilde{G_i})$ and  $H_{i,b}\approx det(\tilde{G_i})$ for $i\in \{1,2\}$, where $\approx_o$ and $\approx$ are opaque bisimulation and bisimulation, respectively. 
		Let $T$ be the largest three-player observer w.r.t.\ $G_1\sync G_2$ and let $T'_i=\langle Q^i_{Y},Q^i_{Z},Q^i_{W},\ACT_i,\ACT_i^{\epsilon}, 
		\Theta_i, \trans^i_{yz},\trans^i_{zz},\trans^i_{zw},\trans^i_{wy},y^i_0\rangle$ be the largest three-player observer w.r.t.\ $det_d(H_{i,ob})$ and $H_{i,b}$  for $i\in \{1,2\}$. Let  $G^T_i=\langle \ACT_{G^T_i},Q_i,\trans^i,Q_i^0,Q_i^m\rangle$ for $i=1,2$ be  
		the transformed automata of $T'_i$ and let 
		$\rho:(\ACT_{G^T_1}\cup\ACT_{G_2^T})\to (\ACT_1\cup \ACT_1^{\epsilon}\cup 
		\Theta_1)\cup (\ACT_2\cup \ACT_2^{\epsilon}\cup \Theta_2)$ be a renaming. We have $[G^T_1\sync G^T_2\trans[s](q_1,q_2)]\ttrans[\tau] [T\trans[\rho(s)]q]$. 
	\end{theorem}
	

	\begin{example}\label{ex:syncTOB}
		Consider the system $\SYSG=\{G_1,G_2\}$ shown in \Fig~\ref{fig:autOFtob}. In the first step of the  compositional approach the system is abstracted by applying opaque observation equivalence, see Example~\ref{ex:abs}. The abstracted system $\tilde{\SYSG}=\{\tilde{G}_1,\tilde{G}_2\}$ is shown in \Fig~\ref{fig:autOFtob}. Next, the three player observer of individual components are built. As explained in Example~\ref{ex:TOBtoAut} the three-player observers of $\tilde{G}_1$ and $\tilde{G}_2$ are  $T'_1$ and $T'_2$, respectively, shown in \Fig~\ref{fig:autOFtob}. Moreover, \Fig~\ref{fig:autOFtob} also shows $G^T_1$ and $G_2^T$, the transformed automata of $T'_1$ and $T'_2$, respectively.  
		The largest monolithic three-player observer w.r.t\ $G_1\sync G_2$ is denoted by $T$ and is shown in \Fig~\ref{fig:syncTOB}. 
		In this particular example, it can be verified that the original and abstracted three-player observers are identical (this need not be true in general); therefore, $T$ in \Fig~\ref{fig:syncTOB} also represents the largest three-player observer w.r.t.\ $\tilde{G_1}\sync \tilde{G_2}$.
		In $T$, we have states: $A=\{(q_0,s_0)\}$, $B=\{(q_0,s_1), (q_0,s_2)\}$, $C=\{(q_1,s_0), (q_2,s_0)\}$, $D=\{(q_1,s_1), (q_2,s_1), (q_2,s_2), (q_1,s_2)\}$ and $E=\{(q_3,s_3)\}$. 
		
		After synchronizing $G^T_1$ with $G^T_2$, we find that there are some transitions in \Fig~\ref{fig:syncTOB}, which do not correspond to  any transition in $G^T_1\sync G^T_2$. For example, no transition in $G^T_1\sync G^T_2$ corresponds to the $zz$ transition of $\beta$ from state $(A, B, \gamma)$ to $(B, B, \gamma)$ in \Fig~\ref{fig:syncTOB}. 

	\end{example}


	\section{FROM ALL EDIT STRUCTURE CALCULATION TO SUPERVISOR SYNTHESIS}\label{sec:sup} 
	
	So far we have shown that in our compositional and abstraction-based approach, individual components can be abstracted and each largest three-player observer w.r.t.\ an abstracted component can be calculated individually. Then we transfer those three-player observers (TPOs) to their automaton forms. After that, we have also shown in Theorem~\ref{pro:trisamerunAfterAbs} that the synchronization of the transformed three-player observers results in a subsystem of the largest monolithic three-player observer up to the renaming of events. 

	Recall that the All Edit Structure (AES) is obtained after pruning deadlocking states from the largest TPO. 
	Here the modular structure of the transformed TPO is kept and the calculation of a ``Modular Edit Structure'' can be done by mapping this problem to a modular  nonblocking supervisory control problem under full observation. 
	As was discussed at the end of Section~\ref{sec:edit-fn}, we pursue an approach to convert the pruning process (from the largest TPO to the AES) to a supervisory control problem. In this setting, the plant is a collection of automata transformed from individual largest TPOs obtained at the end of step (iv) of Algorithm CA-AES. The specification is the automaton form of the edit constraint. 
	The constraint of having up to $n+1$ consecutive erasures can be modeled by a specification automaton with $n$ states where  transitions are labeled by the decision events and all states are marked except the last state, which is a blocking state. After $n$ consecutive event erasures, the next transition of event erasure $\alpha\to \epsilon_\alpha$ leads the specification forward to a blocking state. If the next event is a non-erasure event, it leads the specification back to the initial state, thus resetting the sequence of erasures. Since we have a \emph{modular} representation of the plant, we are able to leverage computationally efficient compositional techniques for modular nonblocking supervisory control problems.

	%
	%
	%

	\begin{definition}\label{def:spec}

		Let $\mathcal{T}=\{T_1,\ldots,T_n\}$ be a three-player observer system where
		$T_i=\langle Q^i_{Y},Q^i_{Z},Q^i_{W},\ACT^i,\ACT_i^{\epsilon}, \Theta^i, \trans^i_{yz},\trans^i_{zz},\trans^i_{zw},\trans^i_{wy},y^i_0\rangle$ and let $\Phi$ be the edit constraint on $\mathcal{T}$ such that there are not $n+1$ 
		consecutive event erasures.  
		Then $K=\langle \ACT_{K}, Q_K, \trans_K, Q_K^0, Q^K_m\rangle$  
		is the automaton form of $\Phi$ where,
		\begin{itemize}
			\item $Q_K=\{x_1,\ldots,x_n\}$
			\item $\trans_K=\bigcup_{1\leq i\leq n-1}\{(x_i,\alpha\to\epsilon_{\alpha},x_{i+1})\mid 
			p\trans[\alpha\to\epsilon]_{zw2}q\ \textnormal{and} \ E(p)=\alpha\}\cup 
			\bigcup_{1\leq i\leq n-2}\{(x_{i+1},\epsilon_\alpha,x_{1})\mid 
			p\trans[\epsilon]_{zw1}q\ \textnormal{and}\ E(p)=\alpha\}\cup 
			\bigcup_{1\leq i\leq n-2}\{(x_{i+1},\alpha_\sigma,x_{1})\mid 
			p\trans[\alpha]_{zz}q\ \textnormal{and}\ E(p)=\sigma\}$ $\cup 
			\{(x_{1},\epsilon_\alpha,x_{1})\mid 
			p\trans[\epsilon]_{zw1}q\ \textnormal{and}\ E(p)=\alpha\}\cup 
			\{(x_{1},\alpha_\sigma,x_{1})\mid 
			p\trans[\alpha]_{zz}q\ \textnormal{and}\ E(p)=\sigma\}$,
			\item $Q_K^0=x_1$
			\item $Q_m^K=\{x_1,\ldots,x_{n-1}\}$
		\end{itemize}	
	\end{definition}

	\begin{figure}
\includegraphics[width=0.85\columnwidth]{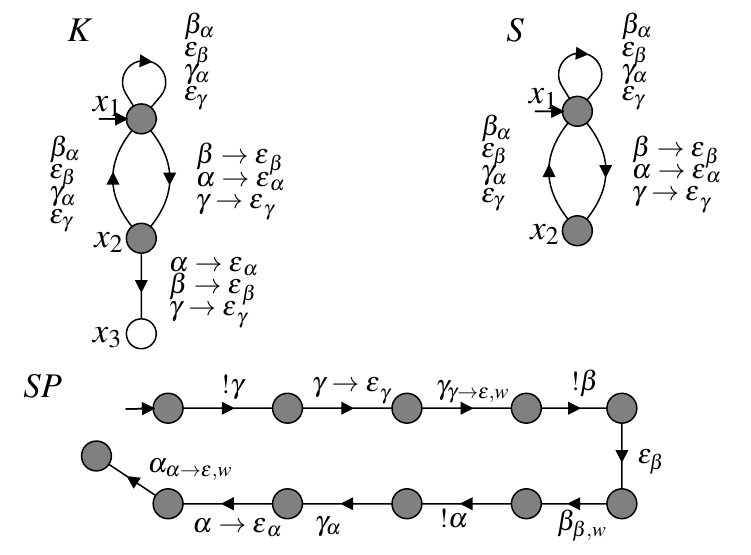}
		\caption{Automaton $K$ is the automaton form of the constraint $\Phi$ in Example~\ref{ex:spec}, $S$ and $SP$ ares the supervisor and the selected path in Example~\ref{ex:final}}.\label{fig:sup}
	\end{figure}

	\begin{example}\label{ex:spec}
		Consider the transformed system $\SYSG^T=\{G_1^T,G_2^T\}$ shown in 
		\Fig~\ref{fig:autOFtob}. Assume the constraint $\Phi$ only allows 
		one erasure. The specification automaton of this constraint 
		is shown in \Fig~\ref{fig:sup} as $K$. Automaton $K$ has three states. As there is no constraint on event insertion, the events related with event insertion just form self-loops at the initial state of $K$. On the other hand, by executing  $\alpha\to\epsilon_\alpha$, $\beta\to\epsilon_\beta$ or $\gamma\to\epsilon_\gamma$ the specification transits from $x_1$ to $x_2$. Next, at $x_2$ if the edit decision is to certain events, then the system goes back 
		to the initial state $x_1$, thus allowing more event erasures since there are no consecutive erasures. However, if another event erasure occurs from $x_2$, the system goes to the blocking state $x_3$. 
	\end{example}
	
	The following theorem establishes that the three-player observer of the system under constraint $\Phi$ and the transformed system synchronized with the specification $K$  have the same runs up to a renaming of the events. 
	
	\begin{theorem}\label{thm:spec}
		Let $G=\langle \Sigma, Q, \rightarrow, Q^0 \rangle$ be a nondeterministic automaton with the set of secret 
		states $Q^S\subseteq Q$.  Let $T'=\langle Q'_{Y},Q'_{Z},Q'_{W},\ACT,\ACT^{\epsilon}, 
		\Theta, \trans'_{yz},\trans'_{zz},\trans'_{zw},\trans'_{wy},y'_0\rangle$ be the largest three-player observer of $G$ under the edit constraint $\Phi$ which prohibits $n+1$ consecutive event erasures. 
		Let $T=\langle Q_{Y},Q_{Z},Q_{W},\ACT,\ACT^{\epsilon}, 
		\Theta, \trans_{yz},\trans_{zz},\trans_{zw},\trans_{wy},y_0\rangle$ be the largest 
		three-player observer w.r.t.\ $G$ when no constraint is considered and let 
		$G^T=\langle\Sigma_{G^T}, Q, \rightarrow, Q^0 \rangle$ be the automaton transformation of $T$. Let $K$ be the specification 
		automaton of $\Phi$. Then $G^T\sync K\trans[s]$ if and only if 
		$T'\trans[\rho(s)]$. 
	\end{theorem}

	\emph{Proof:} 
	Clearly $T'\sqsubseteq T$ and $\ACT_K\subseteq \ACT_{G^T}$. First let $G^T\sync K\trans[s](q_G,q_K)\trans[\sigma](p_G,p_K)$, let 
	$P_K:\ACT_{G^T}\to\ACT_K$ and let $P_E:\ACT_{G^T}\to \bigcup_{\sigma\in\ACT}\{\sigma\to\epsilon_\sigma\in\ACT_K\}$ be a map that removes all the events except event erasures from $\ACT_K$. From $G^T\sync K\trans[s](q_G,q_K)\trans[\sigma](p_G,p_K)$ and $\ACT_K\subseteq \ACT_{G^T}$ and \Defn~\ref{def:synch},  it holds that 
	$G^T\trans[s]q_G\trans[\sigma]p_G$ and $K\trans[P_K(s)]q_K\trans[P_K(\sigma)]p_K$. 
	Now consider three cases: 
	\begin{itemize}
		\item[1)] $q_K=x_i$, $p_K=x_{i+1}$ and $p_K, q_K\in Q_m^K$. Then $|P_E(s)|<n$. This 
		implies $\sigma$ is an erasure event but there are not $n$ consecutive erasures  in $s$, so $T'\trans[\rho(s)]q_T\trans[\rho(\sigma)]p_T$. 
		\item[2)]  $q_K=x_i$, $p_K=x_{i+1}$, $q_K\in Q_m^K$ but $p_K\not\in Q_m^K$. Then $|P_E(s)|=n$ and 
		$K\trans[P_K(s)]q_K\trans[\sigma]p_K$, which implies $G^T\sync K\trans[s\sigma](p_G,p_K)$ and $(p_G,p_K)$ is a blocking state. This further indicates there are $n$ consecutive erasures in $s$,so $T'\trans[\rho(s)]q_T\not\trans[]$. 
		\item[3)] $q_K=x_i$, $p_K=x_{1}$ and $p_K, q_K\in Q_m^K$. This 
		implies $\sigma$ is a non-erasure event, so
		$T'\trans[\rho(s)]q_T\trans[\rho(\sigma)]p_T$. 
	\end{itemize}
	
	Now assume  $T'\trans[\rho(s)]q_T$. This means $T\trans[\rho(s)]q_T$, which implies 
	$G^T\trans[s]q_T$. Consider two cases. 1. $|P_E(s)|<n$. Then 
	$K\trans[P_K(s)]q_K$, which implies $G^T\sync K\trans[s](q_G,q_K)$. 2. $|P_E(s)|=n$. This means 
	$T'\trans[\rho(s)]q_T\not\trans[]$ and $K\trans[P_K(s)]q_K\not\trans[]$, 
	which implies $G^T\sync K\trans[s](q_G,q_K)\not\trans$.\QEDA
	
	The following theorem shows that equivalent states are removed in solving the supervisory control problem to obtain an automaton satisfying the edit constraint and in the pruning process to obtain the AES from the largest TPO. 
	
	
	\def\supervisor{		Let $G=\langle \Sigma, Q, \rightarrow, Q^0 \rangle$ be a nondeterministic automaton with the set of secret 
		states $Q^S\subseteq Q$ and let $\Phi$ be the edit constraint which prohibits $n+1$
		consecutive event erasures. Let
		$T=\langle Q_{Y},Q_{Z},Q_{W},\ACT,\ACT^{\epsilon}, \Theta, \trans_{yz},\trans_{zz},\trans_{zw},\trans_{wy},y_0\rangle$
		be the largest three-player observer w.r.t.\ $G$ without considering the edit constraint and let  $G^T=\langle\Sigma_{G^T}, Q, \rightarrow, Q^0 \rangle$ be the 
		the transformed automaton of $T$. Let 
		$\rho:\ACT_{G^T}\to (\ACT_1\cup \ACT_1^{\epsilon}\cup 
		\Theta_1)$ be a renaming and let $AES$ be the All Edit Structure 
		obtained from $T$.
		Let $S$ be the supremal controllable and 
		nonblocking subautomaton of $G^T$ after considering the specification introduced by $\Phi$. 	
		Then $S\trans[\rho^{-1}(s)]q$ 
		if and only if $AES\trans[\rho(s)]q$.}
	\begin{theorem}\label{thm:supervisor}
    \supervisor
	\end{theorem}
	\emph{Proof:} 
	The pruning process to obtain the AES and the supervisor synthesis procedure are both iterative, which remove states at each iteration.
	We will show by induction that at each iteration a state is removed in $G^T$ by the supervisory control synthesis procedure if and only if the corresponding state is removed by the pruning process from $T$. 
	
	\emph{Base case:} Clearly $G^T$ and $T$ have the same transition relation. 
	
	\emph{Inductive step:} Assume the claim holds for some $n>0$. We let $X^n_G$ be the state space of $G^T$ at the $n$-th iteration of supervisor synthesis process and $X^n_T$ be the state set of $T$ at the $n$-th iteration of pruning process. Then $S\trans[\rho^{-1}(s)]q \Leftrightarrow AES\trans[\rho(s)]q$ holds for all $q\in X^n_G$ ($q\in X^n_T$). 
	Now we need to show that $X^{n+1}_G$ and $X^{n+1}_T$ are also equal. Assume $G^T\trans[s]q$, which implies $T\trans[\rho(s)]q$ based on \Defn~\ref{def:TOBtoAut}.
	
	$(\ttrans)$ First we show that if $q\not\in X^{n+1}_T$, which means $q$ is removed by the pruning process at the $n$-th iteration, so $q\not\in X^{n+1}_G$. Then $q\not\in X^{n+1}_T$ if it is a deadlock state or it is a $Y$ state and there exists $e_o\in \ACT$ such that $q\trans[e_o]z'$, where $z'$ is a deadlock $Z$ state. If $q$ is a deadlock state then based on \Defn~\ref{def:tripartite} it holds that $q$ is either a $W$ state or a $Z$ state. Thus, consider the following three case:
	
	\begin{itemize}
		\item $q$ is a $W$ state. Then $\not\exists e_o\in\ACT$ such that $q\trans[e_o]$. Then based on \Defn~\ref{def:TOBtoAut} it holds that either $q\not\trans[{e_o}_{e_o}]$ or $q\not\trans[{e_o}_{e_o\to\epsilon}]$ in $G^T$ either, which means $q\not\trans$. Thus,  $q$ is a  blocking state in $G^T$ and $q\not\in \Theta^{nonb}(X^n_G)$.
		
		\item $q$ is a $Z$ state.  If $q$ is a deadlock state then $\not\exists\theta\in \Theta$ such that $q\trans[\theta]z'$ or $q\trans[\theta]w$ in $T$. Then based on \Defn~\ref{def:TOBtoAut} it holds that either $q\trans[\theta_{E(q)}]z'$ or $q\trans[\theta_{E(q)}]w$ does not exist in $G^T$, which means $q\not\trans[]$. Thus, $q$ is a blocking state in $G^T$ and $q\not\in \Theta^{nonb}(X^n_G)$.
		
		\item $q$ is a $Y$ state and $q\trans[e_o]z'$, where $z'$ is a deadlock $Z$ state. Then based on \Defn~\ref{def:TOBtoAut} it holds that $q\trans[e_o]z'$ in $G^T$ and $e_o\in\ACT_u$. As it was shown above if $z'$ is a deadlock state in $T$ then $z'$ is also a deadlock state in $G^T$, thus removed by the supervisor synthesis procedure. If $z'$ is removed, i.e, $z'\not\in X_G^n$, then $q\not\in\Theta^{cont}(X^n_G)$.
	\end{itemize}
	Thus, if $q$ is removed in the pruning of $T$, then $q$ is also removed from $G^T$ in supervisor synthesis.

	$(\Leftarrow)$ Now we show that if $q\not\in\Theta^{nonb}(X^n_G)\cap\Theta^{cont}(X^n_G)$, then $q\not\in X^{n+1}_T$ and $q$ needs to be removed from $T$ by the pruning process. If $q\not\in\Theta^{nonb}(X^n_G)\cap\Theta^{cont}(X^n_G)$, there are two cases:
	\begin{itemize}
		\item $q\not\in\Theta^{nonb}(X^n_G)$. Then it holds that $q$ is a blocking state, 
		which means $q\trans[\alpha]$ does not exists in $G^T$. There are three 
		possibilities for $q$, it can be a $Y$, $Z$ or $W$ state. If $q$ is a $Y$ state 
		in $T$ then $q\in Q^m$ in $G$, which means $q\in \hat{\Theta}^{nonb}$, which 
		contradicts the assumption. Thus, $q$ can only be a $W$ or $Z$ state in $T$. In both cases from $q\not\trans[\alpha]$ in $G^T$ and \Defn~\ref{def:TOBtoAut}, it holds that $\not\exists\rho(\alpha)\in \Theta$ such that $q\trans[\rho(\theta)]$ in $T$. This means $q$ is a deadlock state in $T$ and $q\not\in X^{n+1}_T$ 
		
		
		
		\item $q\not\in\Theta^{cont}(X^n_G)$. This means that $q\trans[u]z$ in $G^T$ such that 
		$u\in \ACT_u$ and $z\not\in X^n_G$. Based on \Defn~\ref{def:TOBtoAut}, this means that $q$ is a $Y$ state in $T$ and $z$ is a $Z$ state. It was shown above that if 
		$z\not\in X^n_G$ then $z\not\in X^n_T$, which means $z$ is a deadlock $Z$ state. 
		Thus $q$ will be removed by pruning process, which means $q\not\in X^{n+1}_T$.
	\end{itemize}
	Thus, if $q$ is removed by synthesis from $G^T$ then $q$ is also removed from $T$ by pruning.\QEDA

	Theorem~\ref{thm:supervisor} proves that when it comes to imposing the edit constraint, the pruning process from the largest TPO to the AES removes equivalent states with the synthesis procedure of a supremal supervisor. Hence no information is lost when we apply the  supervisory control approach to enforce the edit constraints and obtain edit functions. 
	This result is essential to show that the transformation of  the TPO to an equivalent automaton and the constraint $\Phi$ to specification $K$ is correctly done in \Defn~\ref{def:TOBtoAutone} and \Defn~\ref{def:spec}, respectively. The next step is to consider the modular representation of the system. In that case, we will use the transformation in \Defn~\ref{def:TOBtoAut}, which results in a set of automata transformations of the individual three-player observers, with necessary self-loops to capture the synchronization among the components. 
	Finally, we combine the results about abstraction and decomposition, which results in Theorem~\ref{thm:complete}.

	
	\begin{theorem}\label{thm:complete}
		Let $\SYSG=\{G_1,\ldots,G_n\}$ be a modular nondeterministic system with  sets of 
		secret states $Q_i^S$. Let $AES$ be the All Edit Structure of 
		$\SYSG$ under constraint $\Phi$. 
		Let $det(G_i)$ and 
		$det_d(G_i)$ be the observer and the desired observer of $G_i$,  
		respectively. 
		Let $\sim_o$ be an opaque observation equivalence 
		on $G_i$ such that $\tilde{G_i}\sim_o G_i$ for $i=1,\cdots,n$. 
		Let $H_{i,ob}\approx_o det(\tilde{G_i})$ and  $H_{i,b}\approx 
		det(\tilde{G_i})$ for $i=1,\cdots,n$, where $\approx_o$ and 
		$\approx$ are opaque bisimulation and bisimulation, respectively. 
		Let	$T'_i$ be the largest three-player observer of  $det_d(H_{i,ob})$ 
		and $H_{i,b}$  for $i=1,\cdots,n$ with the event set $\ACT_{i,T}$.
		Let $G_i^T$ be the transformed 
		automaton of $T'_i$ and $K$ be the automaton specification. Let $P_e:\Omega\to\LANG(\SYSG)$ be an edit projection and $l(\omega)$ be a string generated by run (\Defn~\ref{def:editproj} and \Defn~\ref{def:string-run}).
		Let 
		$\SYSS$ be the least restrictive controllable and nonblocking supervisor calculated 
		from $\{G^T_{1},\ldots,G_{n}^T, K\}$ and let $\rho:(\ACT_{G_1^T}\cup\cdots\cup\ACT_{G_n^T})
		\to (\ACT_{1,T}\cup\cdots\cup\ACT_{n,T})$ be the renaming map. 
		Then [$\forall s\in \LANG(\SYSG)$, $\exists t\in\LANG(\SYSS)$: $P_e(\rho(t))=s$]$\Rightarrow$ [$l(\rho(t))=f_e(s)$ where $f_e\in AES$].
	\end{theorem}
	
	\emph{Proof:} The proof follows directly from Theorems~\ref{thm:supervisor} and~\ref{thm:spec}, in combination with \Thm~\ref{pro:trisamerunAfterAbs}.\QEDA
	

	\Thm~\ref{thm:complete}  essentially shows the proof for all the steps shown in \Fig~\ref{fig:composAlg}. The theorem shows that Algorithm CA-AES correctly synthesizes edit functions for opacity enforcement in a modular form, therefore, the algorithm is sound.  
	It also reveals that the problem of calculating the modular representation of the All Edit Structure can be transformed to synthesizing modular supervisors. 
	The advantage of such a transformation is that we may leverage various existing approaches for calculating a modular supremal nonblocking supervisor in the literature; see, e.g., \cite{feng2008supervisory,schmidt2012efficient, mohajerani2014framework, mohajerani2017compositional}. 
	Therefore, we can obtain a modular representation of the All Edit Structure, which is noticeably efficient to compute. Then we may synchronize individual components in the Modular Edit Structure, which results in a subsystem of the monolithic AES.  
	However, as was pointed out in Section~\ref{sec:AES}, some edit decisions are omitted after the synchronization. In practice, it is usually desired to retain the modular structure and extract an edit function from it, much in the same way as a set of modular supervisors control a plant.
	The extraction process is explained next.
	
\begin{figure*}[ht!]
\includegraphics[scale=0.9]{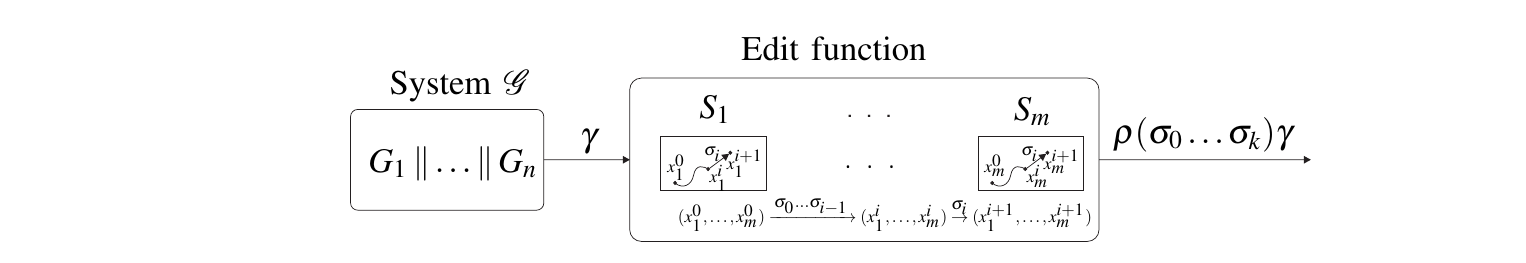}
		\caption{Process of selecting an edit decision from the modular All Edit Structure.}\label{fig:selectingFe}
	\end{figure*}

	Each step of extracting a valid edit decision is described in \Fig~\ref{fig:selectingFe}. Here the edit function is an interface between the system's output and the outside environment. 
	Assume that the system outputs event $\gamma$, then the edit function makes an edit decision for that event and the edited string will be output to the external observers. 
	
	Specifically, this process contains the following steps. $(1)$ when $\gamma$ is received by the edit function, all the components of the Modular Edit Structure are in states that correspond to $Y$ states of the All Edit Structure. 
	$(2)$ At these states, event $\gamma$ is executed and states of all the components in the Modular Edit Structure are updated simultaneously. After the execution of $\gamma$, each component of the Modular Edit Structure is at a state that corresponds to a $Z$ state of the All Edit Structure. $(3)$ Then assume there are multiple transitions defined out of such a current state, we need to select one common transition which corresponds to a specific edit decision and can be viewed as making a control decision from the current state. Note that as the selected transition needs to be accepted by all the components, thus it may happen spontaneously in all the components of the system.  The solution of the modular supervisory control problem guarantees the existence of such a common transition out of the current $Z$ states.
	
	Algorithm CA-AES returns the edited string $\rho(\sigma_0\ldots\sigma_k)$ for event $\gamma$ when every component of the Modular Edit Structure reaches a new state corresponding to a $Y$ state of the AES. 
	At that point, the Modular Edit Structure is ready to process the next event output by the system, and the above steps repeat.  Meanwhile, the algorithm keeps track of the states of the Modular Edit Structure as its components evolve. Based on \Thm~\ref{thm:supervisor}, the edited string $\rho(\sigma_0\ldots\sigma_k)$ is accepted by the monolithic AES. 
	This finally confirms that the extracted edit decision from the Modular Edit Structure corresponds a valid edit decision in the monolithic AES. 
	The above process is illustrated in the following example.

	

	\def\figscale{.42}

	\begin{example}\label{ex:final}
		Consider the nondeterministic system $\SYSG=\{G_1,G_2\}$ shown 
		in \Fig~\ref{fig:autOFtob}. As it was shown in Example~\ref{ex:abs}, the 
		system can be abstracted using opaque observation equivalence. After abstraction, 
		the system becomes deterministic, which means there is no need to calculate the 
		observers of $G_1$ and $G_2$. The largest three-player observers of $\tilde{G_1}$ and $\tilde{G_2}$ are $T'_1$ and $T'_2$, respectively, shown in \Fig~\ref{fig:autOFtob}. 
		Next, the three-player observers are transformed to automata $G_1^T$ and $G_2^T$ shown in \Fig~\ref{fig:autOFtob}, as explained in Example~\ref{ex:TOBtoAut}.

		Assume the user adds an edit constraint such that only one consecutive erasure is 
		allowed as in Example~\ref{ex:spec}. The specification automaton 
		of this constraint is $K$, shown in \Fig~\ref{fig:sup}.
		Due to this constraint, the $Y$ states $(A,D)$ and 
		$(B,E)$ are considered undesired states in $T$, shown in \Fig~\ref{fig:syncTOB} and they should not be reached when we synthesize edit functions. Since $(A, D)$ is not allowed, its successor states $(A, D, \alpha)$, $(A, D, \alpha\rightarrow\epsilon)$ and $(A, E)$ become unreachable from the initial state $(A, A)$. Those three states together with $(A,D)$ and 
		$(B,E)$ are drawn in dashed lines in 
		\Fig~\ref{fig:syncTOB} and are to be removed in the next step. Furthermore, states $(B, D, \alpha \rightarrow \epsilon)$, $(A, C, \beta\rightarrow \epsilon)$ and $(A, B, \gamma\rightarrow \epsilon)$ become deadlocking after $(A,D)$ and 
		$(B,E)$ are removed. They are drawn in dotted lines in 
		\Fig~\ref{fig:syncTOB} and are also to be removed.

		Following the compositional approach with supervisor reduction presented in \cite{su2004supervisor, mohajerani2017compositional}, we calculate a least restrictive and nonblocking supervisor for the transformed automaton, which is shown in \Fig~\ref{fig:sup} as automaton $S$. 
		All the paths accepted by this supervisor represent valid edit decisions. 
		Consider the accepted path $SP$ shown in \Fig~\ref{fig:sup}, it corresponds to an edit function's decisions for string $\gamma\beta\alpha$ such that $f_e(\gamma\beta\alpha)=\gamma(\gamma\to\epsilon)\gamma\beta\epsilon\beta\alpha\gamma(\alpha\to\epsilon)\alpha$, which is shown by thick lines in $T$, \Fig~\ref{fig:syncTOB}.  As is seen, $\rho(SP)=f_e(\gamma\beta\alpha)$. Specifically, when event $\gamma$ is output by the system, $SP$ returns $\gamma({\gamma\to\epsilon}_\gamma)(
		\gamma_{\gamma\to\epsilon,w})$, which means erasing the $\gamma$. Next, event $\beta$ is output by the system and it is unchanged according to $SP$. Finally, $\alpha$ is output by the system and $SP$ returns $\alpha(\gamma_\alpha)(\alpha\to\epsilon_\alpha)(\alpha_{\alpha\to\epsilon,w})$, which means erasing $\alpha$ and inserting $\gamma$. Similarly, we may consider other paths accepted by the supervisor in  \Fig~\ref{fig:sup} to track edit decisions on other strings, then we have a complete picture of how an edit function works. 


	\end{example}

	

	\begin{remark}
		From the result in Theorem~\ref{thm:TOBtoAutSync}, our modular algorithm CA-AES results in fewer edit decisions compared with the monolithic approach in~\cite{ji2019synthesis}, due to the synchronization process in Section~\ref{sec:AES}. This indicates that our method may not be complete in the sense that even if Algorithm CA-AES does not return any modular form edit functions, the monolithic approach may still return valid edit functions. This may be viewed as the tradeoff of reducing computational complexity by the modular method.  \QEDA
	\end{remark}

	\section{CONCLUSION}\label{sec:conclusion} 
	

This paper investigated a compositional and abstraction-based approach to synthesize edit functions for opacity enforcement in a modular setting, given a set of individual systems. The edit functions modify the system's output by inserting and erasing events, under the constraint of limited number of event erasures. 
The Three-Player Observer (TPO) and All Edit Structure (AES) proposed in our prior work were employed here; these discrete structures embed edit functions and reflect the constraints. The monolithic approach first synchronizes all individual systems, then calculates the monolithic AES to obtain edit functions. In contrast, the  compositional approach first exploits the modular structure and builds individual TPOs. Then, it incorporates the edit constraint and calculates the Modular Edit Structure in a nonblocking modular supervisory control manner to obtain edit functions. In addition, we also applied abstraction methods to reduce the state space of the system before opacity enforcement. We showed that the abstraction processes preserve opacity.  Combining system composition and abstraction, we proposed an efficient approach to enforce opacity for complex systems containing multiple components.



\end{document}